\documentclass[11pt]{aastex}
\usepackage[utf8x]{inputenc}
\usepackage{graphicx}
\usepackage{appendix}
\usepackage{ifthen}
\usepackage{longtable}
\usepackage{lscape}

\newcounter{AGNDone}
\def\AGN{\ifthenelse{\equal{\arabic{AGNDone}}{0}}{active galactic nucleus (AGN)\setcounter{AGNDone}{1}}{AGN}}
\def\AGNa{\ifthenelse{\equal{\arabic{AGNDone}}{0}}{active galactic nucleus; AGN\setcounter{AGNDone}{1}}{AGN}}

\newcounter{HMXBDone}
\def\HMXB{\ifthenelse{\equal{\arabic{HMXBDone}}{0}}{high-mass X-ray binary (HMXB)\setcounter{HMXBDone}{1}}{HMXB}}
\def\HMXBs{\ifthenelse{\equal{\arabic{HMXBDone}}{0}}{high-mass X-ray binaries (HMXBs)\setcounter{HMXBDone}{1}}{HMXBs}}

\newcounter{CMBDone}
\def\CMB{\ifthenelse{\equal{\arabic{CMBDone}}{0}}{cosmic microwave background (CMB)\setcounter{CMBDone}{1}}{CMB}}

\newcounter{IMFDone}
\def\IMF{\ifthenelse{\equal{\arabic{IMFDone}}{0}}{initial mass function (IMF)\setcounter{IMFDone}{1}}{IMF}}

\newcounter{IGMDone}
\def\IGM{\ifthenelse{\equal{\arabic{IGMDone}}{0}}{intergalactic medium (IGM)\setcounter{IGMDone}{1}}{IGM}}

\newcounter{ISMDone}
\def\ISM{\ifthenelse{\equal{\arabic{ISMDone}}{0}}{interstellar medium (ISM)\setcounter{ISMDone}{1}}{ISM}}

\newcounter{NFWDone}
\def\NFW{\ifthenelse{\equal{\arabic{NFWDone}}{0}}{Navarro-Frenk-White (NFW)\setcounter{NFWDone}{1}}{NFW}}

\newcounter{SEDDone}
\def\SED{\ifthenelse{\equal{\arabic{SEDDone}}{0}}{spectral energy distribution (SED)\setcounter{SEDDone}{1}}{SED}}

\newcounter{ZAMSDone}
\def\ZAMS{\ifthenelse{\equal{\arabic{ZAMSDone}}{0}}{zero-age main sequence (ZAMS)\setcounter{ZAMSDone}{1}}{ZAMS}}

\begin{document} 

\title{THE ESCAPE FRACTION OF IONIZING RADIATION FROM GALAXIES}

\vspace{1cm}

\author{Andrew Benson$^1$,  Aparna Venkatesan$^2$  \& J. Michael Shull$^3$}

\affil{$^{1}$ Carnegie Observatories, 813 Santa Barbara St, Pasadena, CA 91101}

\affil{$^{2}$ Department of Physics and Astronomy, University of San Francisco, San Francisco, CA 94117}

\affil{$^{3}$ CASA, Department of Astrophysical and Planetary Sciences,  University of Colorado, Boulder, CO 80309}

\email{abenson@obs.carnegiescience.edu, avenkatesan@usfca.edu, michael.shull@colorado.edu}

\begin{abstract}

The escape of ionizing radiation from galaxies plays a critical role in the evolution of gas in galaxies, and the heating and ionization history of the intergalactic medium. We present semi-analytic calculations of the escape fraction of ionizing radiation for both hydrogen and helium from galaxies ranging from primordial systems to disk-type galaxies that are not heavily dust-obscured. We consider variations in the galaxy density profile, source type, location, and spectrum,  and gas overdensity/distribution factors. For sufficiently hard first-light sources, the helium ionization fronts closely track or advance beyond that of hydrogen. Key new results in this work include calculations of the escape fractions for He~I and He~II ionizing radiation, and the impact of partial ionization from X-rays from early AGN or stellar clusters on the escape fractions from galaxy halos.  When factoring in frequency-dependent effects, we find that X-rays play an important role in boosting the escape fractions for both hydrogen and helium, but especially for He~II.  We briefly discuss the implications of these results for recent observations of the He~II reionization epoch at low redshifts, as well as the UV data and emission-line signatures from early galaxies anticipated from future satellite missions.

\end{abstract}

\maketitle

\section{INTRODUCTION}

The escape fraction of ionizing radiation, $f_{\rm esc}$, is a key quantity in the feedback from galaxies on the intergalactic medium (IGM) at all redshifts, $z$, and is an important input parameter in reionization models and cosmological simulations. It is typically defined as the fraction of hydrogen (or when relevant, helium) ionizing photons that escape from a source galaxy to the IGM after accounting for recombinations within the galaxy and its halo. Such radiation profoundly affects the thermal and ionization history of the IGM and is an important factor in the modeling of many astrophysical problems, ranging from metal ionization states in absorber systems to the reionization of the IGM  \citep{loeb_reionization_2001}. The converse problem - that of trapped radiation in galaxies - is also important for understanding the viability of detecting primordial galaxies or Ly$\alpha$ emitters through emission-line and other signatures \citep{tumlinson01,Rhoads:04}.

Theoretical studies \citep{Dove:00,Wood:00,ricotti_feedback_2000,Ricotti:02,gnedin_escape_2008,Wise:09,Razoumov:10,Yajima:11,fernandez_effect_2011} have indicated a wide range of values for $f_{\rm esc}$ for hydrogen, ranging from 1--100\%, beginning with the first detailed studies of the Galaxy's diffuse ionizing gas by \citet{dove_photoionization_1994}. Owing to the complexity and variety of factors that influence photon escape, any single theoretical trend must necessarily be seen as only one aspect of reality.  These factors include galaxy mass and morphology, galaxy redshift, gas density profiles, composition, source luminosities/spectra, source distribution and time evolution within the galaxy, and cloud mass/distribution factors in the interstellar medium (ISM).
Several numerical simulation studies have recently found that $f_{\rm esc}$ increases with decreasing galaxy mass, and with galaxy redshift \citep{gnedin_escape_2008,Shull:12}. Higher mass galaxies will have more gas to ionize (leading to lower $f_{\rm esc}$), but owing to their deeper potential wells, they are also more likely to have higher star formation rates and more collisionally ionized gas in their ISM (leading overall to higher $f_{\rm esc}$, even if the sources are embedded in neutral or molecular gas). Similar complexities arise in redshift studies of $f_{\rm esc}$: in principle, the abundance of lower-mass galaxies and higher star formation rates at high redshifts should lead to higher values of the escape fraction. Some authors find higher values of $f_{\rm esc}$ for dwarf galaxies and Ly$\alpha$ emitters at high redshifts relative to massive galaxies at similar or lower redshifts, but these are complicated by model assumptions underlying these rather different systems \citep{ricotti_feedback_2000,Wood:00,gnedin_escape_2008}.
Direct comparisons with data of low-redshift systems will provide key insights into the many convolved factors, but this has been hindered by a conspicuous lack of numerical simulations of $f_{\rm esc}$  for low-$z$ systems. In addition, values of $f_{\rm esc}$ that exceed 20\% in low-mass galaxies are possibly needed to explain the observational constraints on completing hydrogen reionization by $z \simeq$ 7 \citep{Shull:12,2013arXiv1301.1228R,2013ApJ...763L...7E,2012arXiv1212.4819S,2012arXiv1212.5222M,2012arXiv1212.0860D,Finkelstein:12}. The results of the many theoretical studies, as well as various factors and trends, are summarized comprehensively in \citet{fernandez_effect_2011}.

Observationally, the constraints for $f_{\rm esc}$ have so far been for H alone, focussing (through various techniques) on measuring the escape of Lyman-continuum (LyC) radiation ($h \nu \geq$ 1~Ryd). Here, there has been a gradual approach to some consensus values, and an indication from the past few years of data that $f_{\rm esc, H}$ increases with increasing redshift, confirming one of the theoretical predictions stated above.  
Although observations about a decade ago \citep{Steidel:01} of Lyman continuum emission from Lyman-break galaxies (LBGs)
at  $z \sim$ 3  indicated high values of $f_{\rm esc, H}$ $\ga$ 0.5, more recent analyses of these objects at similar or lower redshifts \citep{Shapley:06,Siana:10,Nestor:11} have derived lower values of $\sim$ 0.1--0.25, with only $\sim$10\% of LBGs and Lyman-$\alpha$ emitters exhibiting large $f_{\rm esc}$ values.  Note that these $f_{\rm esc, H}$ values are a measurement of the {\it relative} escape fraction, the ratio of fractions of escaping ionizing to non-ionizing UV photons, as defined by these authors. There are also indications from measurements of the UV luminosity density of galaxy samples at $z \sim 6$--8 that $f_{\rm esc, H}$  may exceed 50\%  \citep{Finkelstein:12}, in agreement with \citet{Shull:12} and with simulations finding increasing $f_{\rm esc}$ values towards the faint end of the galaxy luminosity function \citep{Ciardi:12}.
At lower redshifts ($z <$ 3), observations have indicated $f_{\rm esc, H}$ $\sim$ 0.01--0.1 from high-mass galaxies \citep{Leitherer:95,Deharveng:01,Heckman:11,grimes_observations_2009}.
Although these fractions appear significantly lower than the $z \sim 3$ values at first, at least some of these systems \citep{Heckman:11} exhibit values of about 30--40\% when recast in terms of the relative escape fraction. Again, it would be valuable to have specific theoretical constraints for comparison, but the field has had very few numerical simulations of $f_{\rm esc}$ at low redshifts.

We note that this relative escape fraction is arguably more comparable to the results of this work (and many others cited above) than the absolute escape fraction. The absolute escape fraction includes, by definition, the effects of any dust obscuration. However, compared to hydrogen photoelectric absorption, dust is usually unimportant for the EUV (LyC).   Our models do not at present consider the effects of dust, as they deal with primordial gas. By considering the relative escape fraction (usually the ratio of escape fractions at 900${\rm \AA}$ and 1500${\rm \AA}$) the effects of dust are mitigated, since the obscuration due to dust should be the same at the two wavelengths modulo the difference in wavelength-dependent opacity\footnote{The difference in opacity between 1500~\AA\ and 1000~\AA\ may typically be a factor of $1.85$ for the standard ISM extinction curve. Future theoretical studies of escape fractions should attempt to incorporate the effects of dust and predict relative escape fractions directly. See \protect\citet{benson_non-uniform_2001} for an example of escape fraction calculations which \emph{do} account for the effects of dust.}. Our main goals here in this work are to model a variety of galaxies that are not heavily obscured by dust, ranging from primordial systems that are more prevalent at high redshifts to the more disk-like galaxies typical at intermediate to low redshifts. 

In contrast to the case of hydrogen, there are no data and detailed theoretical studies of the relevant escape fraction for radiation capable
of ionizing He~II ($h \nu \geq$ 4~Ryd), $f_{\rm esc, He}$. 
In this paper, we derive the values and
geometry of the escape of helium ionizing radiation from the first
galaxies. This quantity can have a profound influence on the thermal and
ionization conditions in the \IGM\, and be an important input for future IGM and  \CMB\
studies, as well as searches for the first stellar clusters. 
We find that the hardness of the source spectrum is an important factor in determining the relative ionization-front (hereafter, ``I-front'') evolution, and we consider both AGN/QSOs as well as  metal-free
stellar populations which, unlike their
low-metallicity counterparts, produce helium ionizing radiation. We also consider the role played by X-rays in the escape of ionizing radiation from early galaxies, particularly for He. We present our model and assumptions in \S2, and the role of frequency-dependent factors in the relative propagation of H and He I-fronts in \S3. We present our results and detailed calculations in \S4, allowing for variations in galaxy density profile, source spectrum, source location, and gas cloud factors, and we conclude in \S5. 

\section{BACKGROUND AND MODEL}\label{sec:model}

Our goal is to compute the emergent spectrum\footnote{Typically, the emergent spectrum is defined as that at infinite distance from the source, but without including radiative transfer through the IGM. In cases where we consider density profiles that extend to infinity (and whose density falls only slowly with distance) we will define the emergent spectrum as that at the virial radius of the dark matter halo associated with the source. Our justification is that the virial radius marks the approximate boundary between galactic and intergalactic environment, and the escape fraction is usually thought of as the fraction of ionizing photons emitted by the source which arrive at the \protect\IGM.}, $S_{R_{\rm vir}}(\nu)$, of ionizing photons emitted by some source (a star cluster or \AGNa) with intrinsic spectrum $S_0(\nu)$ which is embedded within the \ISM\ of a galaxy and/or a more extended gaseous halo. Throughout this work we consider a point source emitting isotropically such that, in absence of absorption, the flux from the source will decline with distance as $1/r^2$. When absorption is included we ignore scattering into and out of the line of sight to the source.

Given the emergent spectrum and the known input spectrum the escape fraction is defined as:
\begin{equation}
 f_{\rm esc} = {\int_{\nu_0}^\infty S_{R_{\rm vir}}(\nu) {\rm d} \nu \over \int_{\nu_0}^\infty S_0(\nu) {\rm d} \nu},
\end{equation}
where $\nu_0$ is the frequency corresponding to the ionization potential for whichever species  is of interest, with $h \nu_0$ = 13.6 eV, 24.6 eV and 54.4 eV for H~I, He~I and He~II respectively. We will consider the gaseous environment of the source to be a mixture of hydrogen and helium, and to have some density distribution $n_i({\bf x})$ for species $i=$ H,He, with the density distributions for hydrogen and helium differing only by a constant proportionality factor appropriate for primordial abundances.

We consider a source located at position ${\bf x}_0$ within this density distribution. Along a direction $\hat{{\bf n}}$ from this source we consider the propagation of the input spectrum through the absorbing gas. We assume that at each distance, $d$, from the source the hydrogen and helium reach photoionization balance at a temperature $T_0$. Given the electron density, $n_{\rm e}$, the equations of photoionization balance imply that 
\begin{eqnarray}
 0 &=& \Sigma_{\rm H\,{\tiny I}} n_{\rm H} (1-x_{\rm H\,{\tiny II}}) - \alpha^{(2)}_{\rm H\,{\tiny II}}(T_0) n_{\rm e} n_{\rm H} x_{\rm H\,{\tiny II}}, \\
 0 &=& \Sigma_{\rm He\,{\tiny I}} n_{\rm He} (1-x_{\rm He\,{\tiny II}}-x_{\rm He\,{\tiny III}}) - \alpha^{(2)}_{\rm He\,{\tiny I}}(T_0) n_{\rm e} n_{\rm He} x_{\rm He\,{\tiny II}} \\
 0 &=& \Sigma_{\rm He\,{\tiny II}} n_{\rm He} x_{\rm He\,{\tiny II}} - \alpha^{(2)}_{\rm He\,{\tiny III}}(T_0) n_{\rm e} n_{\rm He} x_{\rm He\,{\tiny III}},
\end{eqnarray}
where the densities are evaluated at position ${\bf x}_0 + d \hat{{\bf n}}$, and $\Sigma_i$ is
\begin{equation}
 \Sigma_i = \int_{\nu_0}^\infty S(\nu) \sigma_i(\nu) {\rm d} \nu \; .
\end{equation}
Here, $\alpha_i^{(2)}(T)$ is the case-B recombination rate coefficient for ionization state $i$,  $\sigma_i(\nu)$ is the photoionization cross-section for ionization state $i=$H~I, He~I, He~II, and $x_i$ is the fraction of each atomic species in ionization state $i$. Defining $\mathcal{R}_i = \Sigma_i/\alpha_{i+1}^{(2)} n_{\rm e}$, with the convention that
\begin{equation}
 i+1 = \left\{ \begin{array}{ll} {\rm H\,{\tiny II}} & \hbox{ for } i={\rm H\,{\tiny I}} \\  {\rm He\,{\tiny II}} & \hbox{ for } i={\rm He\,{\tiny I}} \\  {\rm He\,{\tiny III}} & \hbox{ for } i={\rm He\,{\tiny II}}, \end{array}\right.
\end{equation}
these equations have solution
\begin{eqnarray}
 x_{\rm H\,{\tiny II}} &=& {\mathcal{R}_{\rm H\,{\tiny II}} \over 1 + \mathcal{R}_{\rm H\,{\tiny II}}}, \\
 x_{\rm He\,{\tiny II}} &=& {\mathcal{R}_{\rm He\,{\tiny II}} \over 1 + \mathcal{R}_{\rm He\,{\tiny II}} + \mathcal{R}_{\rm He\,{\tiny II}} \mathcal{R}_{\rm He\,{\tiny III}}}, \\
 x_{\rm He\,{\tiny III}} &=& {\mathcal{R}_{\rm He\,{\tiny II}} \mathcal{R}_{\rm He\,{\tiny III}} \over 1 + \mathcal{R}_{\rm He\,{\tiny II}} + \mathcal{R}_{\rm He\,{\tiny II}} \mathcal{R}_{\rm He\,{\tiny III}}}.
\end{eqnarray}

We will assume a constant gas temperature of $T_0=10^4$~K throughout, and do not explicitly solve for the temperature. We do not expect the temperature within the I-fronts to be much below $10^4$~K; in fact, photoheated primordial gas has temperatures $\ga$ 20,000~K, which has even been measured in the Ly$\alpha$ forest \citep{Becker:11}.
For the $10^8M_\odot$ halo at $z=10$ that we will consider in \S\ref{sec:profile} the virial temperature is $\sim 1.1 \times 10^4$~K \citep{Donahue:91,Becker:11}.

The ionization balance depends on the electron density, which can be found directly from the ionization states:
\begin{equation}
 n_{\rm e} = n_{\rm H} x_{\rm H\,{\tiny II}} + n_{\rm He} (x_{\rm He\,{\tiny II}} + 2 x_{\rm He\,{\tiny III}}).
\end{equation}
Therefore, we make an initial guess at the ionization states to allow us to compute $n_{\rm e}$ and then iteratively update the ionization fractions using the above equations until we obtain a converged solution for $n_{\rm e}$. After finding the ionization states we compute the optical depth to photoionization across a small increment in distance $\delta d$ using
\begin{equation}
 \delta \tau = \delta d \sum_i \int_{\nu_{0,i}}^\infty n_{i^\prime} x_i \sigma_i(\nu) {\rm d} \nu,
\end{equation}
where $i^\prime=$H for $i=$H~I and $i^\prime=$He for $i=$He~I or He~II. The step in distance, $\delta d$ is chosen to ensure $\delta \tau \ll 1$ and $\delta d \ll r_{\rm s}$ where $r_{\rm s}$ is the characteristic length scale of the density profile being considered. The spectrum is then updated using
\begin{equation}
 S(\nu) \rightarrow S(\nu) {\exp(-\delta \tau) \over (1+\delta d/d)^2},
\end{equation}
where the term in the numerator on the right hand side accounts for absorption of photons by the H and He, and the term in the denominator accounts for the $1/r^2$ reduction in flux with distance from the source. We repeat this process until a sufficiently large distance is reached that the spectrum is no longer changing significantly or, when considering source embedded in dark matter halos, the \IGM\ (i.e. the halo virial radius) is reached.  We then compute the escape fractions, $f_{i,{\rm esc}}(\hat{{\bf n}})$, as defined above. The mean escape fraction, averaged over all directions, is then
\begin{equation}
 \langle f_{i,{\rm esc}} \rangle = {1\over 4\pi} \int f_{i,{\rm esc}}(\hat{{\bf n}}) {\rm d}\Omega.
\end{equation}

\subsection{Inclusion of clouds}

To include clouds (i.e. regions of enhanced density) in our calculations we follow \citet{fernandez_effect_2011}. Clouds are defined by three quantities: their radius ($r_{\rm c}$), their volume filling factor ($f_{\rm c}$), and the cloud overdensity, $C$, the ratio of density in clouds to non-cloud regions (referred to as the ``clumping factor''  in \citealt{fernandez_effect_2011}). 

Along each direction, $\hat{{\bf n}}$, we construct a cylindrical volume with radius equal to the cloud radius such that any cloud whose center lies within this cylinder will intersect the line of sight along which the escape fraction is being computed. We find the mean number of clouds in this cylinder based on their radius
and volume filling factor, $\langle N_{\rm c} \rangle = f_{\rm c} \pi r_{\rm c}^2 d_{\rm max}/(4 \pi/3) r_{\rm c}^3$, where $d_{\rm max}$ is the maximum distance from the source to be considered. The actual number of clouds in the cylinder is then drawn from a Poisson distribution with this mean. Each cloud is located randomly within the cylinder and the section of the line of sight that the cloud intersects is determined.

In propagating the spectrum through this cloudy medium we now choose the steps in distance $\delta d$ such that cloud boundaries always precisely coincide with a step. Therefore any given step
is either in a cloud or not in a cloud (i.e. a step is never partly in a cloud and partly not). In cloud regions the gas density is increased appropriately\footnote{It is possible for clouds to overlap. In such cases we still increase the density only by a factor of $C$ relative to the background, not by a factor of $2C$ for example. This overlap means that the actual volume filling factor, $f_{\rm c}^\prime$, will be slightly smaller than the input filling factor, $f_{\rm c}$: $f_{\rm c}^\prime = 1-\exp(-f_{\rm c})$ on average. For the cases we consider, $f_{\rm c}=0.2$, so $f_{\rm c}^\prime=0.18$, a small difference.} (see \citealt{fernandez_effect_2011}). The emergent spectrum of photons then accounts for the effects of clouds along the line of sight.

\section{THE EVOLUTION OF HELIUM IONIZATION FRONTS} 

\subsection{The Critical Spectral Index}

We begin by calculating the critical spectral index at which the rate of H
and He photoionizations become equal. For a source spectrum whose specific
luminosity (erg s$^{-1}$ Hz$^{-1}$) goes as $L_\nu \propto \nu^{-\alpha}$,
we find that the ratio of photoionization rates per atom for He and H, $X_{\rm c}$, is:
\\
\begin{equation}
X_{\rm c} \equiv \frac{Q({\rm He\,{\tiny II}})/n_{\rm He}}{Q({\rm H\,{\tiny I}})/n_{\rm H}} =
\left[ \frac{\int_{{\rm 4 Ryd}/h}^\infty d\nu L_\nu/(h \nu)}{\int_{{\rm 1
      Ryd}/h}^\infty d\nu
  L_\nu/(h \nu)} \right] \frac{n_{\rm H}}{n_{\rm He}}.
 \label{eq:heEv}
\end{equation}

\noindent
The primordial helium mass fraction $Y_{\rm He} \approx 0.25$ \citep{peimbert_revised_2007,aver_new_2010}, and the currently permitted range of $\Omega_{\rm b} h^2 = 0.02255 \pm 0.00054 $ \citep{komatsu_seven-year_2011}. This leads to a range of values for the relative helium fraction by number,
$y = n_{\rm He}/n_{\rm H}$ = 0.0789--0.0823. We find that $X_{\rm c}$ is unity when:
\begin{equation}
4^{-\alpha_{\rm crit}} = 0.08, \; {\rm or} \; \alpha_{\rm crit} = - \frac{\ln (y)}{\ln (4)} \simeq 1.82 \pm 0.03,
\label{eq:16}
\end{equation}
if we assume $y =  0.080\pm0.003$.  
Therefore, in the case of a source with $\alpha \la 1.82$, the helium I-fronts may begin to overtake or coincide with the H I-fronts; such a hard spectral index may already be seen in $z \sim$ 0.03--1.45 QSO spectra \citep{shull_hst-cos_2012} as we discuss below.  For a source spectrum with
multiple indices, e.g., $\alpha_1$ between $\nu_1$ and $\nu_2$, and $\alpha_2$
beyond $\nu_2$, equation~(\ref{eq:heEv}) is unity when:

\begin{equation}
\frac{\nu_2^{\alpha_2}}{\nu_1^{\alpha_1}} \times \frac{\alpha_1}{\alpha_2} \times 4^{-\alpha_2} = 0.08.
\end{equation}

\begin{figure}[!t]
\begin{center}
\includegraphics[width=150mm]{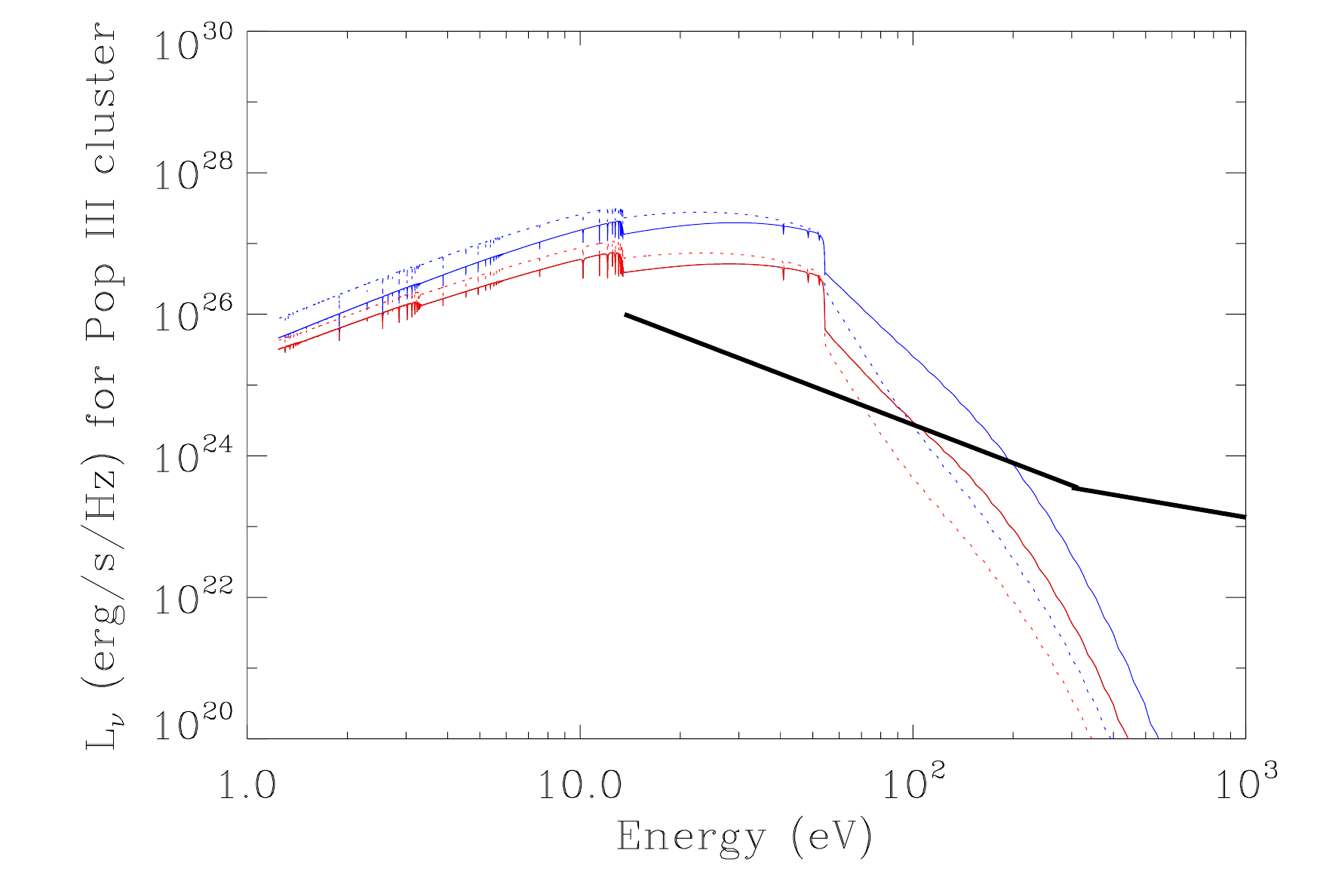}
\end{center}
\caption{The specific luminosity as a function of energy of a fiducial
  $10^6$ $M_\odot$ metal-free stellar cluster, based on \citet{tumlinson_nucleosynthesis_2004}. 
  Red and blue lines represent 1-100 $M_\odot$ and 10-140 $M_\odot$ IMFs, with solid and dashed 
  lines in each case denoting the values on the zero-age main sequence (ZAMS) and at 2 Myr.  Note the
  relative flatness (hardness) of the spectrum between the H~I and He~II
  ionizing thresholds at 13.6 eV and 54.4 eV respectively, and the rapid
  decline of the He ionizing flux with time. For comparison, a typical QSO
  spectrum (arbitrarily normalized) with slope $\alpha$ = 1.8 and 0.8 in the respective energy
  ranges 13.6--300 eV and $\ga$ 300 eV is also shown.}
 \label{fig:spectra}
\end{figure}

To extend this estimate to the case of metal-free stars, we now consider
three cases involving metal-free IMFs in the mass ranges 1--100 $M_\odot$
\citep{tumlinson_cosmological_2003}, 10--140 $M_\odot$ (\citet{tumlinson_nucleosynthesis_2004}; 
hereafter TVS04), and $\ga$ 100 $M_\odot$ \citep{bromm_generic_2001}, all with a Salpeter slope 
unless noted otherwise.  The first represents a roughly present-day \IMF, the second the best-case
\IMF\ that matches the current data on nucleosynthesis and reionization, and
the third, an \IMF\  consisting of very massive stars (VMSs). Such stars can
lead to pair-instability SNe in the mass range 140-260 $M_\odot$ and may
dominate in primordial gas that lacks metal coolants. Thus 140 $M_\odot$
divides stellar masses into two regimes that have similar radiative but
different nucleosynthetic yield properties. For the first two of these
IMFs, we display the specific intensity of a fiducial $10^6 M_\odot$
zero-metal (Pop III) cluster as a function of energy in Figure~\ref{fig:spectra}. In each
of these cases, two curves corresponding to the \ZAMS\ and at times of 2 Myr
are shown; for all the curves, we observe the relative hardness of the
cluster's radiation between 1 Ryd and 4 Ryd. A numerical evaluation of
Equation~(\ref{eq:heEv}) yields a value $X_{\rm c} \simeq$ 0.24 (0.07) for the \ZAMS\ (2 Myr) 1-100
$M_\odot$ \IMF, and $X_{\rm c} \simeq$ 0.43 (0.11) for the \ZAMS\ (2 Myr) 10-140
$M_\odot$ \IMF.

The fact that $X_{\rm c}$ never exceeds unity implies that the relative rate of He to H photoionizations remains below one, and therefore the He I-front evolution cannot exceed that of H for these IMFs. This is mostly due to the sharp fall off of
the intensity beyond the He ionization threshold. For comparison, Figure 1 also shows the spectrum of a 
quasar (arbitrarily normalized) with slope $\alpha$ = 1.8 and 0.8 in the  energy ranges 13.6-300 eV and 
$\ga$ 300 eV respectively \citep{Venkatesan:01}. Such a source would clearly meet the criterion in 
Equation~(\ref{eq:16}), and the He I-fronts could ``keep up" with those associated with H.   AGN with 
sufficiently hard spectral indices are already observed in COS spectra of $z \sim$ 0.03--1.45 QSOs \citep{shull_hst-cos_2012} where composite spectral slopes $\alpha$ = 1.4--1.5 are observed at 
energies 1.0--1.5 Ryd. This is a harder spectral slope than the $\alpha_{\rm crit}$ derived earlier, if it continues 
to the 4 Ryd edge for He~II reionization. However, this is within the range of the predicted critical spectral index when frequency-dependent effects in the source spectrum are factored in, as well as the range detected observationally in low-$z$ AGN with COS. We now show this in the next subsection.

\subsection{Frequency-dependent Effects in the Source Spectrum}

In order to compare the values of  $f_{\rm esc}$ for H and He for any general galaxy density profile, we begin with 
Equation~(12) of  \citet{dove_photoionization_1994},
\begin{equation}
f_{\rm esc}  = 1 - \frac{4 \pi \alpha_{\rm B}}{S_0} \int_{0}^\infty n_{\rm e} n_{\rm H} r^2 {\rm d}r   \; .  
\end{equation} 
In order to assess the impact of frequency dependence in the input
spectrum, we replace $S_0$ with a frequency-dependent function for the source photon luminosity, $S_0 (\nu/\nu_0)^{-(\alpha + 1)}$. 
Thus, the first part of the second term in the equation above becomes:
\begin{equation}
\frac{4 \pi \alpha_{\rm B}}{\int d\nu S_0 (\nu/ \nu_0)^{-(\alpha + 1)}}   \; .  
\end{equation}
The constant $S_0$ has to be evaluated through integration from a
threshold energy, e.g., from $h \nu_{\rm H}$ = 1 Ryd or $h \nu_{\rm
  He}$ = 4 Ryd, up to infinite energy, and by setting the integral equal
to the H-ionizing (or He-ionizing) photon flux in units of photons
s$^{-1}$. If we evaluate the above integral between the ionization
threshold energy of interest up to
energies $h \nu_1 = A (h \nu_0$), where $\nu_1$ is a constant multiple
of $\nu_0$, then the integral reduces to (switching variables to
$\nu/\nu_0$):

\begin{equation}
\nu_0 S_0 \int_{1}^{A}  d\left(\frac{\nu}{\nu_0}\right) \, \left(\frac{\nu}{\nu_0}\right)^{-(\alpha + 1)} = \nu_0 S_0 [1 - A^{-\alpha}], {\; \rm for \; H}  \; .  
\end{equation}

\begin{equation}
\nu_0 S_0 \int_{4}^{A}  d\left(\frac{\nu}{\nu_0}\right) \left(\frac{\nu}{\nu_0}\right)^{-(\alpha + 1)} = \nu_0 S_0 [4^{-\alpha} - A^{-\alpha}], {\; \rm for \; He}  \; .  
\end{equation}
For H~I, $h \nu_0$ = 1 Ryd, whereas for He~II, $h \nu_0$ = 4 Ryd. Note
that the constant $S_0$ through its normalization is always
proportional to the product $\alpha S_{\rm 0, H}/[\nu_0 (1 - A^{-\alpha})]$
for H, and to the product $\alpha S_{\rm 0, He}/[\nu_0 (4^{-\alpha} -
  A^{-\alpha})]$ for He. When integrating up to infinity (i.e., $A = \infty$), the last term on the right-hand side simply goes to zero.  Therefore, the analytic expressions for the escape fractions for H and He with source frequency dependence will respectively have the general form:

\begin{equation}
f_{\rm esc,H} = 1 - \frac{4 \pi \alpha_{\rm B, H}}{\alpha S_{\rm 0, H}} n_{\rm e, c} n_{\rm H, c}  \left[\int_{0}^\infty \rho(r)^2 r^2 {\rm d}r \right] \; ,
\end{equation}
and
\begin{equation}
f_{\rm esc,He} = 1 - \frac{4 \pi \alpha_{\rm B, He}}{\alpha S_{\rm 0, He}} n_{\rm e, c} n_{\rm He, c} \left[\int_{0}^\infty \rho(r)^2 r^2 {\rm d}r \right] \; ,
\end{equation}
where $\rho(r)$ is the galaxy density profile. We explore several cases of $\rho(r)$ below in Section 4, but focus here on the relative values of $f_{\rm esc}$ for H and He from frequency-dependent effects in the source spectrum. To compare with the calculation of the critical spectral index of $\alpha \sim1.8$ derived in Section 3.1 when H and He photoionizations became equal, we set the second terms on the RHS of the two above equations equal to each other. This leads to:
\begin{equation}
4^{-\alpha} = 0.08 \left(\frac{\alpha_{\rm B, He}}{\alpha_{\rm B, H}}\right), \; {\rm or} \; \alpha_{\rm eff} \simeq 0.5 \; ,
\end{equation}
for the standard values for the recombination coefficients and assuming that $n({\rm He})/n({\rm H}) = 0.08$.  Here, $\alpha_{\rm eff}$ is the critical source spectral index after gas filtering and transmission effects are taken into account. This value of $\alpha_{\rm eff} \sim$ 0.5 is lower than the $\alpha_{\rm crit} \sim$ 1.82 derived in Section 3.1 which did not account for these effects. Both these values, however, encompass the range found in UV spectroscopic studies of low-$z$ AGN \citep{zheng_composite_1997,shull_hst-cos_2012,syphers_hst/cos_2012}.


\section{RESULTS}

Our model contains a number of parameters which control both the source spectrum and the distribution of matter around that source. In this section we explore a representative range of parameters. A summary of the results obtained can be found in Table~\ref{tb:ResultsSummary}.

\begin{landscape}
\begin{table}
\begin{tabular}{llrrllrrrrrrrrr}
\hline
{\bf Profile} & {\bf Source}            & \boldmath{$M$}         & \boldmath{$d$}     & {\bf X?} & {\bf C?}     & \boldmath{$Q_{\rm 0,HI}$} & \boldmath{$Q_{\rm 0,HeI}$} & \boldmath{$Q_{\rm 0,HeII}$} & \boldmath{$f_{\rm esc,HI}$} & \boldmath{$f_{\rm esc,HeI}$} & \boldmath{$f_{\rm esc,HeII}$} & \boldmath{$\theta_{\rm c}$} \\
        &                   & \boldmath{$[M_\odot]$} & {\bf [kpc]} &    &        &  {\bf [s\boldmath{$^{-1}$}]}              &  [s\boldmath{$^{-1}$}]               & {\b [s\boldmath{$^{-1}$}]}                 &  {\bf [\%]}            &  {\bf [\%]}             &  {\bf [\%]}     & \boldmath{$[^\circ]$}         \\
\hline
$\beta$ &               AGN & $1.0(4)$    &  0    & N  &      N & $1.17(50)$     & $6.84(49)$      & $4.10(49)$       &   1              &   2               &   0                &   --             \\
$\beta$ &               AGN & $1.0(4)$    &  0    & N  &      Y & $1.17(50)$     & $6.84(49)$      & $4.10(49)$       &   0              &   0               &   0                &   --             \\
$\beta$ &               AGN & $1.0(4)$    &  0    & Y  &      Y & $1.17(50)$     & $6.84(49)$      & $4.10(49)$       &  10              &  18               &  30                &   --             \\
$\beta$ &               AGN & $1.0(4)$    & 10    & N  &      Y & $1.17(50)$     & $6.84(49)$      & $4.10(49)$       &   0              &   0               &   0                &   --             \\
$\beta$ &               AGN & $1.0(4)$    & 10    & Y  &      Y & $1.17(50)$     & $6.84(49)$      & $4.10(49)$       &  12              &  20               &  33                &   --             \\
$\beta$ &               AGN & $1.0(6)$    &  0    & N  &      N & $2.76(52)$     & $1.14(52)$      & $4.28(51)$       & 100              & 100               &  99                &   --             \\
$\beta$ &               AGN & $1.0(6)$    &  0    & N  &      Y & $2.76(52)$     & $1.14(52)$      & $4.28(51)$       &  99              &  99               &  97                &   --             \\
$\beta$ &               AGN & $1.0(6)$    &  0    & Y  &      Y & $2.76(52)$     & $1.14(52)$      & $4.28(51)$       &  99              &  99               &  98                &   --             \\
$\beta$ &               AGN & $1.0(6)$    & 10    & N  &      Y & $2.76(52)$     & $1.14(52)$      & $4.28(51)$       &  99              &  99               &  97                &   --             \\
$\beta$ &               AGN & $1.0(6)$    & 10    & Y  &      Y & $2.76(52)$     & $1.14(52)$      & $4.28(51)$       &  99              &  99               &  98                &   --             \\
$\beta$ &            PopIII & $1.0(4)$    &  0    & N  &      N & $3.81(51)$     & $2.30(51)$      & $1.30(50)$       &  97              &  98               &  71                &   --             \\
$\beta$ &            PopIII & $1.0(4)$    &  0    & N  &      Y & $3.81(51)$     & $2.30(51)$      & $1.30(50)$       &  93              &  95               &  32                &   --             \\
$\beta$ &            PopIII & $1.0(4)$    &  0    & Y  &      Y & $3.81(51)$     & $2.30(51)$      & $1.30(50)$       &  93              &  95               &  34                &   --             \\
$\beta$ &            PopIII & $1.0(4)$    & 10    & N  &      Y & $3.81(51)$     & $2.30(51)$      & $1.30(50)$       &  93              &  95               &  38                &   --             \\
$\beta$ &            PopIII & $1.0(4)$    & 10    & Y  &      Y & $3.81(51)$     & $2.30(51)$      & $1.30(50)$       &  93              &  95               &  39                &   --             \\
$\beta$ &            PopIII & $1.0(6)$    &  0    & N  &      N & $3.81(53)$     & $2.30(53)$      & $1.30(52)$       & 100              & 100               & 100                &   --             \\
$\beta$ &            PopIII & $1.0(6)$    &  0    & N  &      Y & $3.81(53)$     & $2.30(53)$      & $1.30(52)$       & 100              & 100               &  99                &   --             \\
$\beta$ &            PopIII & $1.0(6)$    &  0    & Y  &      Y & $3.81(53)$     & $2.30(53)$      & $1.30(52)$       & 100              & 100               &  99                &   --             \\
$\beta$ &            PopIII & $1.0(6)$    & 10    & N  &      Y & $3.81(53)$     & $2.30(53)$      & $1.30(52)$       & 100              & 100               &  99                &   --             \\
$\beta$ &            PopIII & $1.0(6)$    & 10    & Y  &      Y & $3.81(53)$     & $2.30(53)$      & $1.30(52)$       & 100              & 100               &  99                &   --             \\
$\beta$ &  PopIII$^\dagger$ & $1.0(4)$    & 10    & Y  &      Y & $3.81(51)$     & $2.30(51)$      & $1.30(50)$       &  93              &  95               &  39                &   --             \\
$\beta$ &  PopIII$^\dagger$ & $1.0(6)$    & 10    & Y  &      Y & $3.81(53)$     & $2.30(53)$      & $1.30(52)$       & 100              & 100               &  99                &   --             \\
D\&S    &               AGN & $1.0(4)$    & --    & N  &      Y & $1.17(50)$     & $6.84(49)$      & $4.10(49)$       &  31              &  35               &  23                & 45.2             \\
D\&S    &               AGN & $1.0(4)$    & --    & Y  &      Y & $1.17(50)$     & $6.84(49)$      & $4.10(49)$       &  40              &  48               &  51                & 45.2             \\
\hline
\end{tabular}
\caption{Escape fractions for all models shown in this work. ``Profile'' specifies the functional form of the density profile, while ``Source'' specifies the nature of the photon source, $M$ is the mass of that source, $d$ its distance from the center of the density profile. Columns ``X?'' and ``C?'' indicate whether X-rays and clouds are included respectively (see text for standard cloud model parameters). For models using the \cite{dove_photoionization_1994} profile (labelled ``D\&S''), the critical angle, $\theta_{\rm c}$, is shown in the final column. Also shown are the input ionizing fluxes, $Q_0$, in units of photons per second. Numbers in parentheses indicate the exponent, e.g. $1.0(4)$ means $1.0 \times 10^4$. $^*$ Model ``BB1'' is a blackbody source with a temperature of $7.94\times 10^4$~K, model ``BB2'' is a blackbody source with a temperature of $1.15\times 10^5$~K. $^\dagger$ Model spectrum includes a contribution from \protect\HMXBs. $^\ddagger$ Model does not include helium. $^\P$ Model has cloud radii of 3~pc.}
\label{tb:ResultsSummary}
\end{table}
\end{landscape}
\begin{landscape}
\begin{table}
\begin{tabular}{llrrllrrrrrrrrr}
\hline
{\bf Profile} & {\bf Source}            & \boldmath{$M$}         & \boldmath{$d$}     & {\bf X?} & {\bf C?}     & \boldmath{$Q_{\rm 0,HI}$} & \boldmath{$Q_{\rm 0,HeI}$} & \boldmath{$Q_{\rm 0,HeII}$} & \boldmath{$f_{\rm esc,HI}$} & \boldmath{$f_{\rm esc,HeI}$} & \boldmath{$f_{\rm esc,HeII}$} & \boldmath{$\theta_{\rm c}$} \\
        &                   & \boldmath{$[M_\odot]$} & {\bf [kpc]} &    &        &  {\bf [s\boldmath{$^{-1}$}]}              &  [s\boldmath{$^{-1}$}]               & {\b [s\boldmath{$^{-1}$}]}                 &  {\bf [\%]}            &  {\bf [\%]}             &  {\bf [\%]}     & \boldmath{$[^\circ]$}         \\
\hline
D\&S    &               AGN & $1.0(5)$    & --    & Y  &      Y & $1.86(51)$     & $9.08(50)$      & $4.36(50)$       &  73              &  76               &  73                & 73.7             \\
D\&S    &               AGN & $1.0(6)$    & --    & N  &      Y & $2.76(52)$     & $1.14(52)$      & $4.28(51)$       &  88              &  89               &  82                & 83.4             \\
D\&S    &               AGN & $1.0(6)$    & --    & Y  &      Y & $2.76(52)$     & $1.14(52)$      & $4.28(51)$       &  89              &  89               &  86                & 83.4             \\
D\&S    & BB1$^*$$\ddagger$ &       --    & --    & Y  &      N & $1.40(51)$     & $6.03(50)$      & $2.81(49)$       &  80              &  84               &  86                & 72.0             \\
D\&S    & BB1$^*$$\ddagger$ &       --    & --    & Y  &      N & $1.40(52)$     & $6.03(51)$      & $2.81(50)$       &  91              &  92               &  94                & 81.8             \\
D\&S    &           BB1$^*$ &       --    & --    & Y  &      N & $1.40(51)$     & $6.03(50)$      & $2.81(49)$       &  78              &  79               &  39                & 72.0             \\
D\&S    &           BB1$^*$ &       --    & --    & Y  &      N & $1.40(52)$     & $6.03(51)$      & $2.81(50)$       &  90              &  90               &  70                & 81.8             \\
D\&S    &           BB1$^*$ &       --    & --    & Y  &      Y & $1.40(52)$     & $6.03(51)$      & $2.81(50)$       &  86              &  87               &  61                & 81.8             \\
D\&S    &           BB1$^*$ &       --    & --    & Y  & Y$^\P$ & $1.40(52)$     & $6.03(51)$      & $2.81(50)$       &  86              &  87               &  61                & 81.8             \\
D\&S    &           BB2$^*$ &       --    & --    & Y  &      Y & $1.16(53)$     & $6.96(52)$      & $9.73(51)$       &  93              &  94               &  88                & 85.9             \\
D\&S    &           BB2$^*$ &       --    & --    & Y  &      Y & $1.16(55)$     & $6.96(54)$      & $9.73(53)$       &  99              &  99               &  97                & 89.1             \\
D\&S    &            PopIII & $1.0(4)$    & --    & N  &      Y & $3.81(51)$     & $2.30(51)$      & $1.30(50)$       &  79              &  81               &  54                & 77.2             \\
D\&S    &            PopIII & $1.0(4)$    & --    & Y  &      Y & $3.81(51)$     & $2.30(51)$      & $1.30(50)$       &  79              &  81               &  54                & 77.2             \\
D\&S    &            PopIII & $1.0(5)$    & --    & Y  &      Y & $3.81(52)$     & $2.30(52)$      & $1.30(51)$       &  90              &  91               &  78                & 84.1             \\
D\&S    &            PopIII & $1.0(6)$    & --    & N  &      Y & $3.81(53)$     & $2.30(53)$      & $1.30(52)$       &  95              &  96               &  89                & 87.3             \\
D\&S    &            PopIII & $1.0(6)$    & --    & Y  &      Y & $3.81(53)$     & $2.30(53)$      & $1.30(52)$       &  95              &  96               &  90                & 87.3             \\
NFW     &               AGN & $1.0(4)$    &  0    & N  &      Y & $1.17(50)$     & $6.84(49)$      & $4.10(49)$       &   0              &   0               &   0                &   --             \\
NFW     &               AGN & $1.0(4)$    &  0    & Y  &      Y & $1.17(50)$     & $6.84(49)$      & $4.10(49)$       &   5              &   9               &  15                &   --             \\
NFW     &               AGN & $1.0(4)$    & 10    & N  &      Y & $1.17(50)$     & $6.84(49)$      & $4.10(49)$       &   0              &   0               &   0                &   --             \\
NFW     &               AGN & $1.0(4)$    & 10    & Y  &      Y & $1.17(50)$     & $6.84(49)$      & $4.10(49)$       &   9              &  15               &  25                &   --             \\
NFW     &               AGN & $1.0(6)$    &  0    & N  &      Y & $2.76(52)$     & $1.14(52)$      & $4.28(51)$       &  94              &  94               &  79                &   --             \\
NFW     &               AGN & $1.0(6)$    &  0    & Y  &      Y & $2.76(52)$     & $1.14(52)$      & $4.28(51)$       &  94              &  95               &  88                &   --             \\
NFW     &               AGN & $1.0(6)$    & 10    & N  &      Y & $2.76(52)$     & $1.14(52)$      & $4.28(51)$       &  94              &  95               &  85                &   --             \\
NFW     &               AGN & $1.0(6)$    & 10    & Y  &      Y & $2.76(52)$     & $1.14(52)$      & $4.28(51)$       &  95              &  95               &  90                &   --             \\
\hline
\end{tabular}
\addtocounter{table}{-1}
\caption{Escape fractions for all models shown in this work. ``Profile'' specifies the functional form of the density profile, while ``Source'' specifies the nature of the photon source, $M$ is the mass of that source, $d$ its distance from the center of the density profile. Columns ``X?'' and ``C?'' indicate whether X-rays and clouds are included respectively (see text for standard cloud model parameters). For models using the \cite{dove_photoionization_1994} profile (labelled ``D\&S''), the critical angle, $\theta_{\rm c}$, is shown in the final column. Also shown are the input ionizing fluxes, $Q_0$, in units of photons per second. Numbers in parentheses indicate the exponent, e.g. $1.0(4)$ means $1.0 \times 10^4$. $^*$ Model ``BB1'' is a blackbody source with a temperature of $7.94\times 10^4$~K, model ``BB2'' is a blackbody source with a temperature of $1.15\times 10^5$~K. $^\dagger$ Model spectrum includes a contribution from \protect\HMXBs. $^\ddagger$ Model does not include helium. $^\P$ Model has cloud radii of 3~pc.}
\end{table}
\end{landscape}
\begin{landscape}
\begin{table}
\begin{tabular}{llrrllrrrrrrrrr}
\hline
{\bf Profile} & {\bf Source}            & \boldmath{$M$}         & \boldmath{$d$}     & {\bf X?} & {\bf C?}     & \boldmath{$Q_{\rm 0,HI}$} & \boldmath{$Q_{\rm 0,HeI}$} & \boldmath{$Q_{\rm 0,HeII}$} & \boldmath{$f_{\rm esc,HI}$} & \boldmath{$f_{\rm esc,HeI}$} & \boldmath{$f_{\rm esc,HeII}$} & \boldmath{$\theta_{\rm c}$} \\
        &                   & \boldmath{$[M_\odot]$} & {\bf [kpc]} &    &        &  {\bf [s\boldmath{$^{-1}$}]}              &  [s\boldmath{$^{-1}$}]               & {\b [s\boldmath{$^{-1}$}]}                 &  {\bf [\%]}            &  {\bf [\%]}             &  {\bf [\%]}     & \boldmath{$[^\circ]$}         \\
\hline
NFW     &            PopIII & $1.0(4)$    &  0    & N  &      Y & $3.81(51)$     & $2.30(51)$      & $1.30(50)$       &  67              &  79               &   0                &   --             \\
NFW     &            PopIII & $1.0(4)$    &  0    & Y  &      Y & $3.81(51)$     & $2.30(51)$      & $1.30(50)$       &  67              &  79               &   0                &   --             \\
NFW     &            PopIII & $1.0(4)$    & 10    & N  &      Y & $3.81(51)$     & $2.30(51)$      & $1.30(50)$       &  78              &  82               &  22                &   --             \\
NFW     &            PopIII & $1.0(4)$    & 10    & Y  &      Y & $3.81(51)$     & $2.30(51)$      & $1.30(50)$       &  78              &  82               &  22                &   --             \\
NFW     &            PopIII & $1.0(6)$    &  0    & N  &      Y & $3.81(53)$     & $2.30(53)$      & $1.30(52)$       & 100              & 100               &  96                &   --             \\
NFW     &            PopIII & $1.0(6)$    &  0    & Y  &      Y & $3.81(53)$     & $2.30(53)$      & $1.30(52)$       & 100              & 100               &  96                &   --             \\
NFW     &            PopIII & $1.0(6)$    & 10    & N  &      Y & $3.81(53)$     & $2.30(53)$      & $1.30(52)$       & 100              & 100               &  96                &   --             \\
NFW     &            PopIII & $1.0(6)$    & 10    & Y  &      Y & $3.81(53)$     & $2.30(53)$      & $1.30(52)$       & 100              & 100               &  96                &   --             \\
\hline
\end{tabular}
\addtocounter{table}{-1}
\caption{Escape fractions for all models shown in this work. ``Profile'' specifies the functional form of the density profile, while ``Source'' specifies the nature of the photon source, $M$ is the mass of that source, $d$ its distance from the center of the density profile. Columns ``X?'' and ``C?'' indicate whether X-rays and clouds are included respectively (see text for standard cloud model parameters). For models using the \cite{dove_photoionization_1994} profile (labelled ``D\&S''), the critical angle, $\theta_{\rm c}$, is shown in the final column. Also shown are the input ionizing fluxes, $Q_0$, in units of photons per second. Numbers in parentheses indicate the exponent, e.g. $1.0(4)$ means $1.0 \times 10^4$. $^*$ Model ``BB1'' is a blackbody source with a temperature of $7.94\times 10^4$~K, model ``BB2'' is a blackbody source with a temperature of $1.15\times 10^5$~K. $^\dagger$ Model spectrum includes a contribution from \protect\HMXBs. $^\ddagger$ Model does not include helium. $^\P$ Model has cloud radii of 3~pc.}
\label{table:1}
\end{table}
\end{landscape}

\subsection{Gaussian Distribution Model}\label{sec:DS}

We begin by considering the simple Gaussian density distribution model of \citet{dove_photoionization_1994}. We use this paper (rather than the many others cited in earlier sections) for general comparison, as its analytic results are straightforward to reproduce, at least for a pure hydrogen case. It also permits easy comparison with our results which include a number of variations in galaxy density profiles and source spectra, as well as the specific new scenarios including helium and X-rays.  Adopting the infinite slab model given in Equation~(7) of \cite{dove_photoionization_1994}, we assume that:
\begin{equation}
 n_{\rm H}({\bf x}) = n_0 \exp(-[Z/h]^2/2),
\end{equation}
where $n_0=0.312$ cm$^{-3}$, the Gaussian scale height $h=0.215$~kpc, and $Z$ is the vertical distance above the midplane of the slab. The density profile for helium is the same, scaled by the cosmic helium to hydrogen ratio. We adopt a blackbody input spectrum with a temperature of $10^{4.9}$~K normalized such that the production rate of H~I ionizing photons is comparable to that in the models of Pop III stars that we will consider later. Specifically, we consider a model (model ``BB1'' in Table~\ref{tb:ResultsSummary}) with ionizing photon rates ($Q$) for H~I, He~I and He~II of, respectively, $1.4 \times 10^{51}$ s$^{-1}$,  $6.1 \times 10^{50}$ s$^{-1}$, and $2.9 \times 10^{49}$ s$^{-1}$, and a second model (model ``BB2'' in Table~\ref{tb:ResultsSummary}) with the $Q$ values simply scaled up by a factor of 10.  In this model, there are relatively few X-rays present, although we consider scenarios below with AGN that do explicitly include them. We have also considered the contribution from high-mass X-ray binaries (HMXBs) which may form nearly simultaneously with the Pop~III stars. We adopt the model of \cite{mirabel_stellar_2011} to specify the normalization of the HMXB spectrum. Specifically, their Equation~(5) gives the luminosity in the 2--10~keV range per unit star formation rate. We find that the inclusion of HMXBs makes negligible difference to the escape fractions and so do not consider them further in this work.

\begin{figure}
 \begin{center}
\begin{tabular}{c}
 \includegraphics[width=150mm]{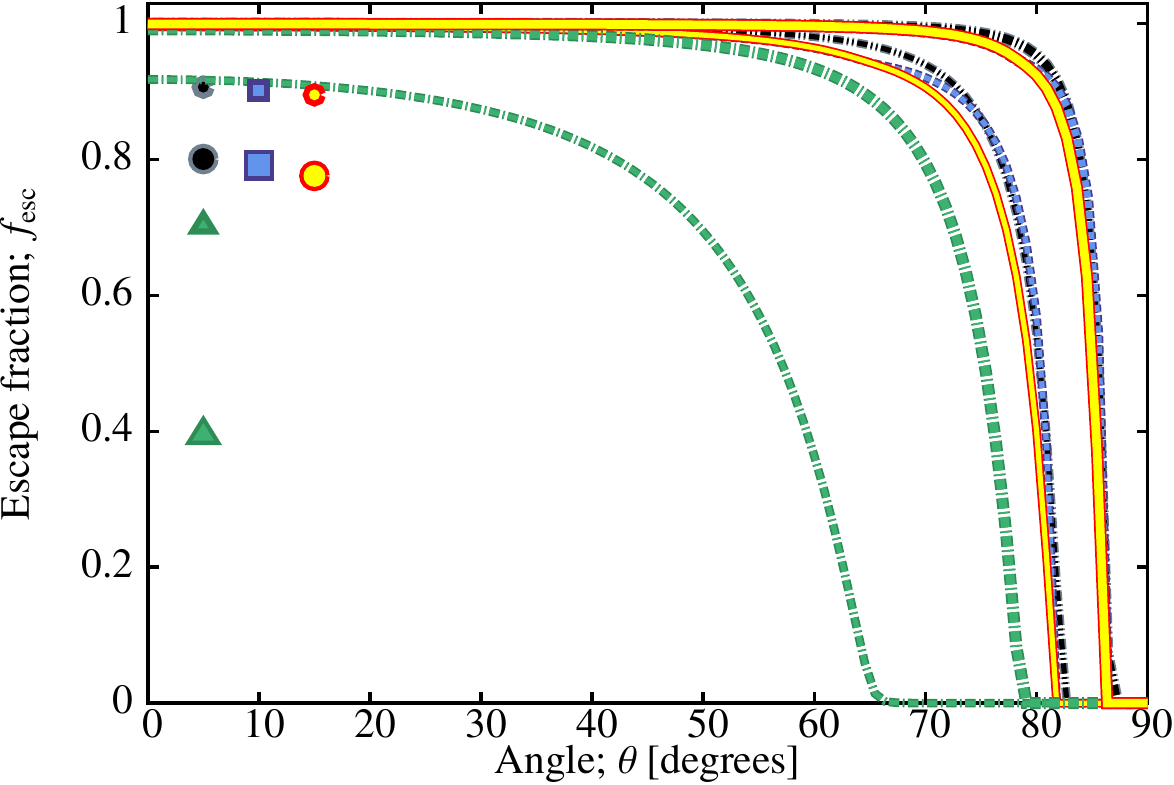}  \vspace{-50mm}\\
\includegraphics[width=40mm]{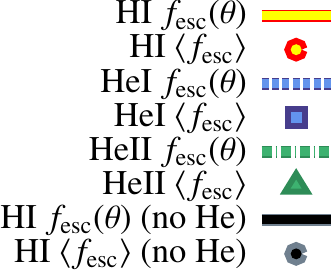}\\
\end{tabular}
 \end{center}
\vspace{+10mm} \caption{The escape fraction as a function of angle, $\theta$, in the \protect\citet{dove_photoionization_1994} Gaussian model. Thick and thin lines correspond to $10^4M_\odot$ and $10^5M_\odot$ Pop~III clusters respectively. Yellow (solid), blue (dashed) and green (dot-dashed) correspond to H~I, He~I and He~II respectively. The net escape fractions (i.e. the escape fraction averaged over solid angle) are shown by symbols: yellow (circles), blue (squares) and green (triangles) correspond to H~I, He~I and He~II respectively.  Small and large symbols corresponding to $10^4M_\odot$ and $10^5M_\odot$ Pop~III clusters respectively. For comparison black (double dot-dashed) lines show the escape fraction for H~I when no helium is included in the calculation, with the black circle showing the corresponding angle-averaged escape fraction.}
 \label{fig:DSangle}
\end{figure}

For these $Q$-values, the analytic formula in Equation~(14) from \citet{dove_photoionization_1994} for the critical opening angle, $\theta_{\rm c}$, of the cone of escaping H-ionizing radiation yields values of approximately 72$^\circ$ and 82$^\circ$ for the $10^4$ $M_\odot$ and $10^5$ $M_\odot$ Pop~III clusters respectively. This is in excellent agreement with the results of the semi-analytic calculation described in \S\ref{sec:model}, as shown in Fig.~\ref{fig:DSangle}. Specifically, there is close agreement with the angles at which $f_{\rm esc}$ begins to drop significantly below unity, i.e., the ionizing radiation is being contained\footnote{\protect\citet{dove_photoionization_1994} use a Str\"omgren argument to determine whether any photons along a given line of sight can escape. As such, their escape fraction as a function of angle is a monotonically declining function, going from $1$ to $0$ at the critical angle. In our calculation we account for the non-zero mean free path of photons through the gas. This results in a smooth dependence of escape fraction on angle, making a comparison with \protect\citet{dove_photoionization_1994} non-trivial. Given this caveat, we find the agreement between our results and the critical angle calculation of \protect\citet{dove_photoionization_1994} to be acceptable.}. The black lines in Fig.~\ref{fig:DSangle} show the escape fraction for H~II when we remove helium from our calculation -- clearly the presence of helium causes only a small change in the escape of hydrogen-ionizing photons in this particular case.

Repeating this calculation for He~II and He~III (with their respective recombination coefficients and $Q$-values as stated earlier), we find that for He~II (He~III), $\theta_c$ $\sim$ 77$^\circ$ (11$^\circ$)  and 84$^\circ$ (63$^\circ$) for the $10^4$ $M_\odot$ and $10^5$ $M_\odot$ Pop~III clusters.  This assumes that all of the helium (for $y_{\rm He}$ = 0.08) is either in He~II or He~III for these $\theta_c$ values. Comparing with the curves in Fig.~\ref{fig:DSangle}, we see that, similar to hydrogen, the $\theta_c$ for each case corresponds to where the curves depart from their maximum value, although this is less well-defined for He~III. This perhaps reveals the limitations of extending the \citet{dove_photoionization_1994} analysis to helium, especially when we do not account for (in this simple estimate) the absorption of helium-ionizing photons by hydrogen. We do see that the critical angle for escaping He~II ionizing radiation is less than that for H~I, as predicted by the analytic results earlier.

\begin{figure}
 \begin{center}
\begin{tabular}{c}
 \includegraphics[width=150mm]{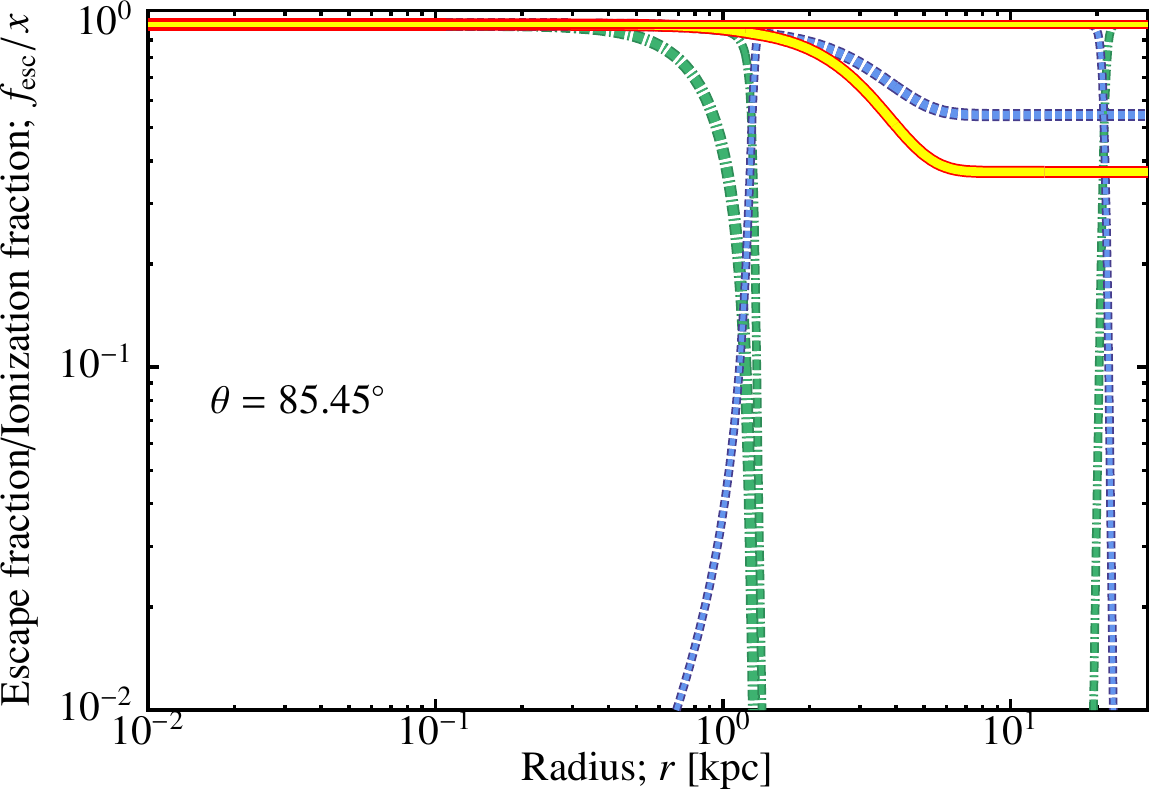}\vspace{-48mm}\\
\hspace{-20mm}\includegraphics[width=40mm]{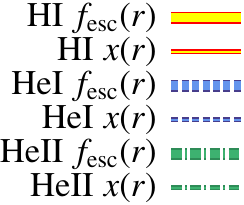} \\
\end{tabular}
 \end{center}
\vspace{+10mm}  \caption{The escape fraction (thick lines) and ionization state (thin lines) as a function of distance from the source at $\theta=85.45^\circ$ (the angle at which the H~I escape fraction drops precipitously---see Fig.~\protect\ref{fig:DSangle}), in the Dove \& Shull (1994) Gaussian model for a $10^5M_\odot$ Pop~III cluster. Yellow (solid), blue (dashed) and green (dot-dashed) correspond to H~I, He~I and He~II respectively.}
 \label{fig:DSradial}
\end{figure}

In Figure~\ref{fig:DSradial}, we show the escape and ionization fractions as a function of radial distance from the source, evaluated at the critical angle of $\theta \sim 85^\circ$.
Note that in Figure~\ref{fig:DSradial}, the He~II fraction rises again at large radii ($\ga$ 10 kpc). The density drops so rapidly at these large radii that even the weak radiation field that has escaped to this distance is able to ionize the gas. While interesting in its own regard, this is mostly irrelevant for the escape fraction calculation since the density (and optical depth) are extremely low at these large radii, especially given our constant temperature assumption. (At these low densities, cooling rates will be very low, driving the gas to higher temperatures at which collisional ionization becomes relevant.)

Last, we comment on the geometry of ionized helium versus hydrogen bubbles in the context of \citet{dove_photoionization_1994}.
Beginning with their Equation~(12), we can recast it  as, $f_{\rm esc, i}(\theta) = 1 - W_i$, where $W_i \propto (\alpha_i^{(2)} n_{\rm i})/Q_i$, Here, $i$ corresponds to either H or He, and the constant of proportionality is related to the product of the gas distribution and the electron density (both of which are the same for H and He). The product $(\alpha_i^{(2)} n_{\rm i})$ is already set for each species, and is typically less for He~II and He~III relative to H~II, whereas for typical Pop~III clusters, $Q_H$ significantly exceeds $Q_{He}$. Therefore, in principle, $W_{\rm He} > W_{\rm H}$, and $f_{\rm esc, He} < f_{\rm esc, H}$. Additionally,  extending Equation~(14) in \citet{dove_photoionization_1994}, 
\begin{equation}
\frac{\cos(\theta_{\rm c, H})}{\cos(\theta_{\rm c, He})} = \left[ \frac{W_{\rm H}}{W_{\rm He}} \right]^{1/3} \; ,
\end{equation}
which will typically have values less than 1, therefore implying that $\theta_{\rm c, He} < \theta_{\rm c, H}$. Thus, the escape of He-ionizing radiation will occur through narrower opening angles than for H-ionizing radiation. These broad conclusions will hold in the absence of any coupling of their ionization equilibria through, e.g., X-rays or otherwise.

\subsection{Blackbody source and cloud factors}

\begin{figure}
 \begin{center}
 \begin{tabular}{cc}
\multicolumn{2}{c}{$C=10$; $r_{\rm c}=30$~pc; $f_{\rm c}=20$\%}\\
 \includegraphics[width=80mm]{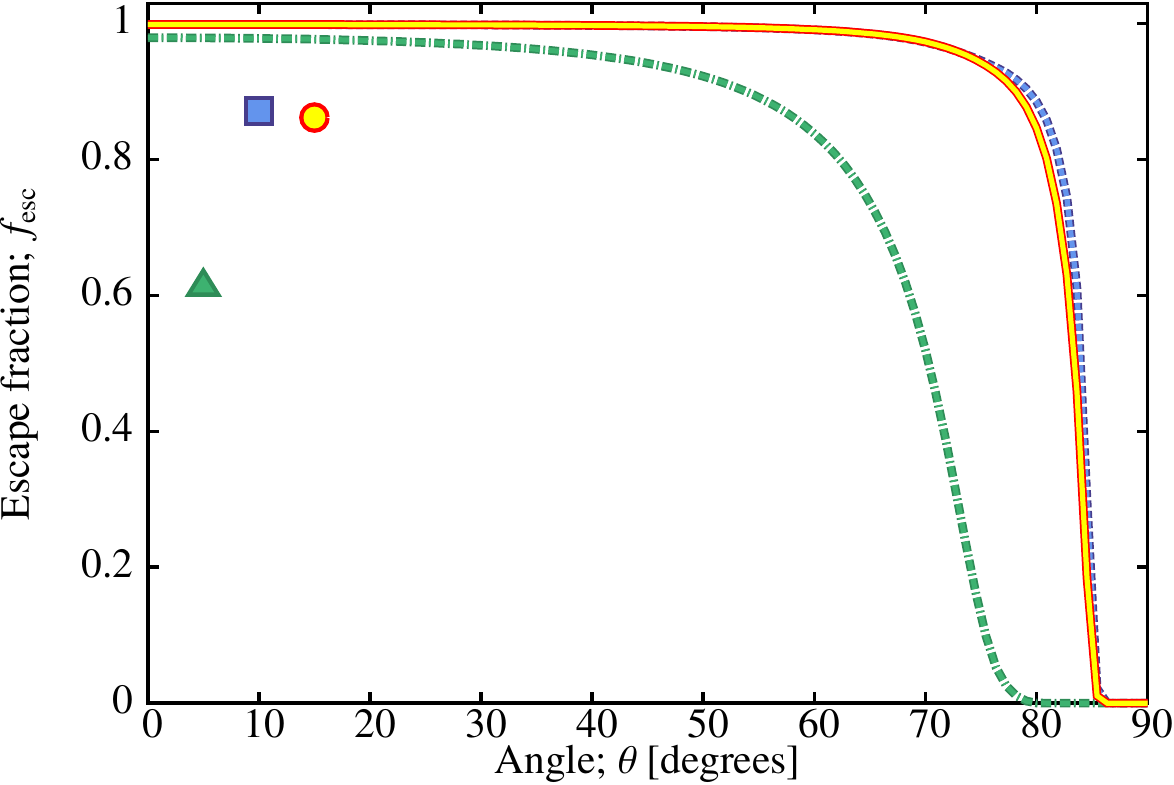} &
 \includegraphics[width=80mm]{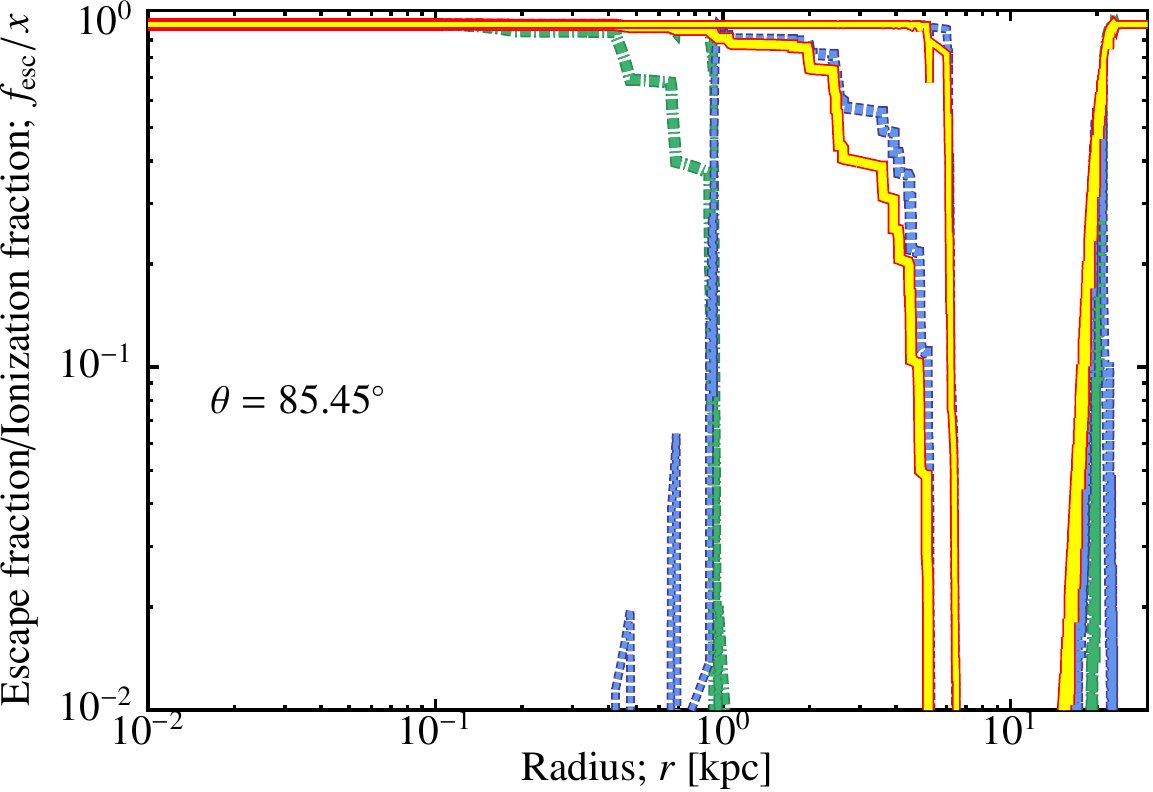}\vspace{-25mm}\\
\hspace{-20mm}\includegraphics[width=20mm]{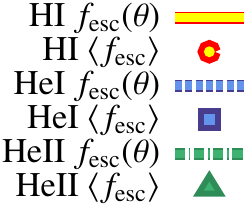} &
\hspace{-20mm}\includegraphics[width=20mm]{legend4.pdf} \\
 \end{tabular}
 \end{center}
\vspace{+10mm}  \caption{\emph{Left panel:} The escape fraction as a function of angle, $\theta$, for a single realization of the \protect\citet{dove_photoionization_1994} Gaussian model for a $10^5M_\odot$ Pop~III cluster with clouds of radius 30~pc, filling factor 20\% and cloud overdensity factor of $C=10$. Yellow (solid), blue (dashed) and green (dot-dashed) correspond to H~I, He~I and He~II respectively. The net escape fractions (i.e. the escape fraction averaged over solid angle) are shown by symbols. Yellow (circles), blue (squares) and green (triangles) correspond to H~I, He~I and He~II respectively. \emph{Right panel:} The escape fraction (thick lines) and ionization state (thin lines) as a function of radial distance from the source at $\theta=85.45^\circ$ for the same model.}
\label{fig:DS:single}
\end{figure}

We next consider the effects of adding in clouds to our model. The left panel of Figure~\ref{fig:DS:single} shows the escape fraction as a function of angle for the $10^5M_\odot$ Pop~III cluster model of \S\ref{sec:DS} but now with a single realization\footnote{Note that we use the same radial distribution of clouds for each angle, $\theta$, so there are no fluctuations in escape fraction with angle. This is unrealistic of course, but this case serves only to illustrate the effects of clouds. Our primary interest is in the mean escape fraction averaged over many realizations of the cloud population, for which the lack of angular fluctuations in our calculations is irrelevant.} of a population of 30~pc radius spherical clouds with a volume filling factor of 20\% and a cloud overdensity factor of $C=10$. The denser cloud regions stay more neutral and provide greater optical depth to the ionizing photons. Therefore, the escape fractions are lower than the case with no clouds.

The right panel of Figure~\ref{fig:DS:single} shows the radial variation of escape fractions and ionization fractions for the same model. Large fluctuations in ionization state are apparent---these are the difference between in-cloud and out-of-cloud points. The escape fraction lines show corresponding steps as the photons get strongly depleted when they encounter a cloud.

\begin{figure}
 \begin{center}
 \begin{tabular}{cc}
$C=10$; $r_{\rm c}=30$~pc; $f_{\rm c}=20$\% & $C=10$; $r_{\rm c}=3$~pc; $f_{\rm c}=20$\% \\
 \includegraphics[width=80mm]{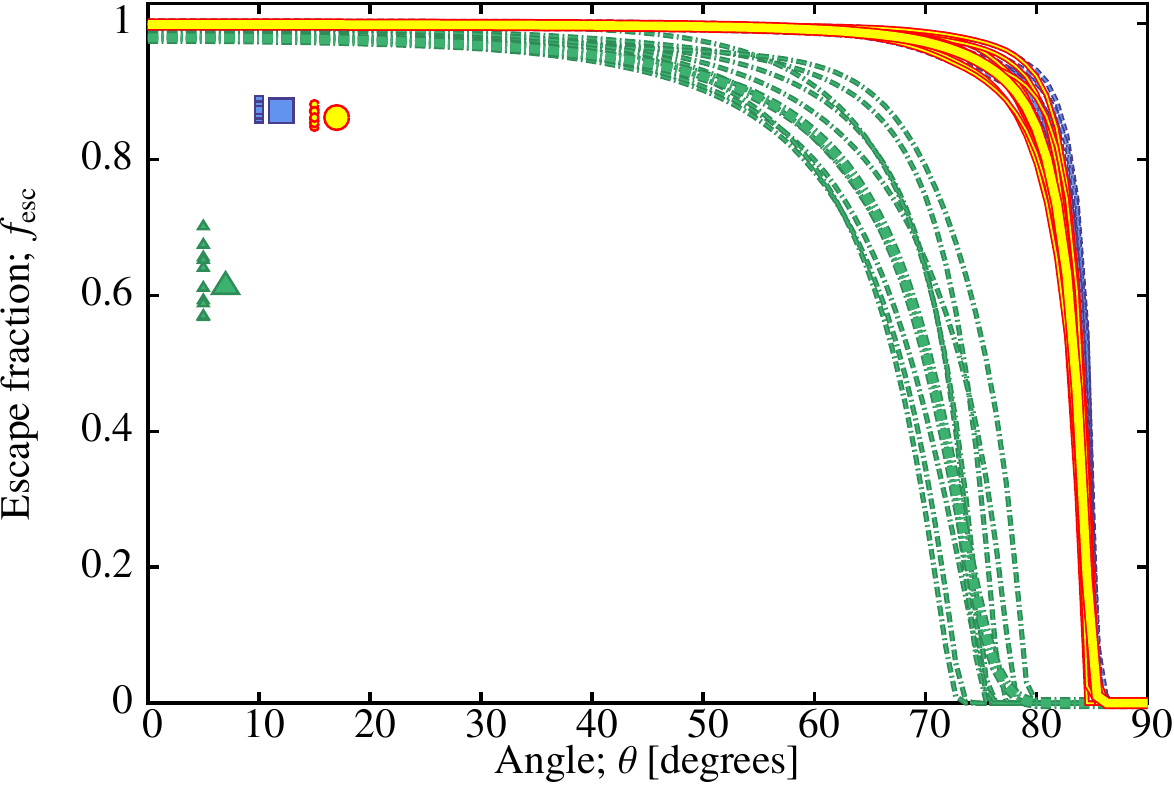} & 
 \includegraphics[width=80mm]{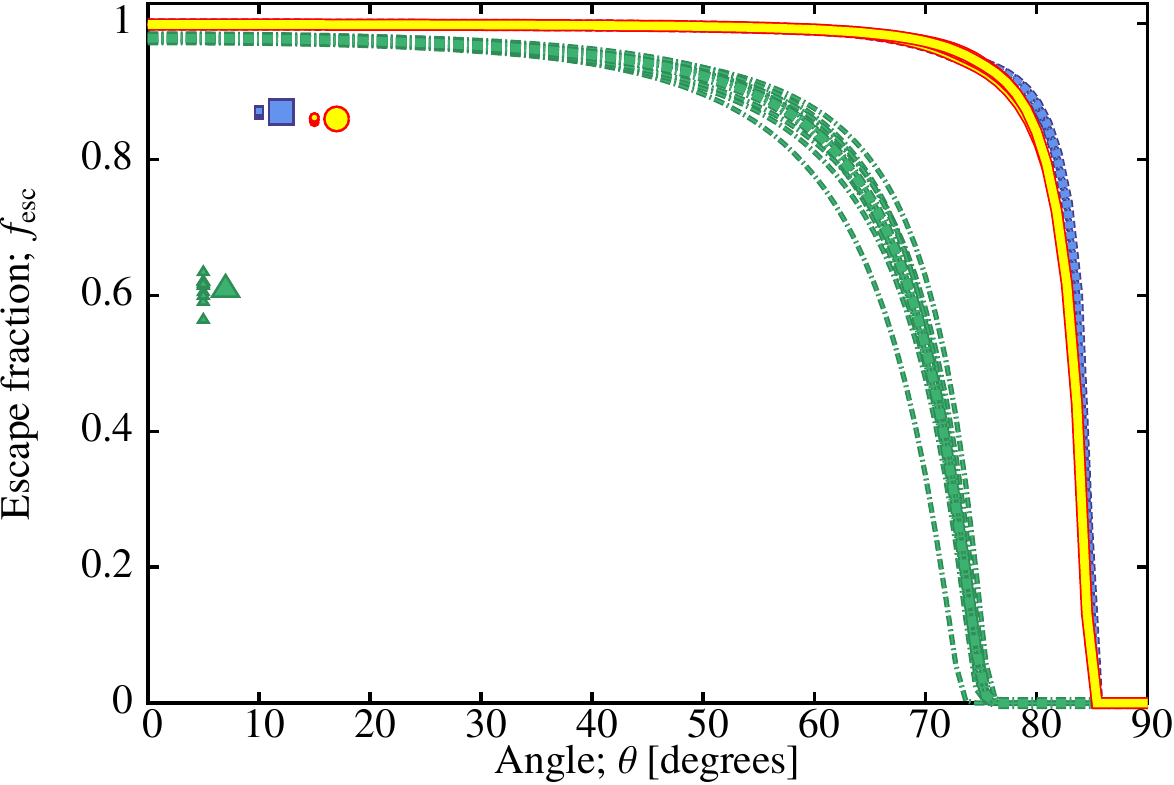}\vspace{-30mm}\\
\hspace{-10mm}\includegraphics[width=20mm]{legend3.pdf} \\
 \end{tabular}
 \end{center}
 \vspace{+10mm} \caption{\emph{Left panel:} The escape fraction as a function of angle, $\theta$, for ten realizations (thin lines) and the average over many realizations (thick lines) of the \protect\citet{dove_photoionization_1994} Gaussian model for a $10^5M_\odot$ Pop~III cluster with clouds of radius 30~pc, filling factor 20\% and cloud overdensity factor of $C=10$. Same notation as Figure~\ref{fig:DS:single}. Small symbols show mean escape fractions for individual realizations while large symbols show the mean escape fraction averaged over a large number of realizations. \emph{Right panel:} The equivalent result for the same model but with clouds of radius 3~pc.}
\label{fig:DSangle:multiple}
\end{figure}

The left-panel of Figure~\ref{fig:DSangle:multiple} shows the escape fraction as a function of angle for ten realizations of the cloud population (thin lines) and the average over a much larger number of realizations (thick lines). Once again, we use clouds of 30~pc radius, with a volume filling factor of 20\% and $C=10$. There is significant variation in escape fraction between realizations---the angle-averaged escape fraction for He~III varies from about 55\% to 70\% with the mean being close to 60\%.

To explore the effects of cloud size, the right panel of Figure~\ref{fig:DSangle:multiple} shows the escape fraction for the same model but now using clouds of 3~pc radius. As expected, there is
much less variation in the escape fraction when using these smaller clouds---since the volume filling factor remains the same, the number of clouds is increased by a factor of 1000, reducing the fluctuation from realization to realization. The mean escape fractions are not significantly changed relative to the case of 30~pc clouds however. Note also that helium is especially sensitive to the presence of clouds, with individual realizations departing significantly from the average.

\subsection{Variations in Source Spectrum}

We now consider \AGN\ and Pop~III starburst cases in the \citet{dove_photoionization_1994} Milky Way background density profile described earlier, with a cloud filling factor of 20\%, cloud radius 30~pc and cloud overdensity factor of $C=10$. Additionally, we now include cases with and without X-rays (in the latter of which the source spectrum is cut off above 120~eV). The \AGN\ spectrum is modeled from \citet{hopkins_observational_2007}---this gives the unabsorbed, intrinsic \SED\ of \AGN\ as a function of their bolometric luminosity which we in turn compute from an assumed black hole mass, Eddington limit and accretion efficiency (we assume black holes to be radiating at 10\% of their Eddington luminosity). For this \AGN\ source, we find $(Q_{\rm H~I},Q_{\rm He~I},Q_{\rm He~II})=1.17,0.68,0.41\times10^{50}$~photons~s$^{-1}$ for a $10^4M_\odot$ black hole and $(Q_{\rm H~I},Q_{\rm He~I},Q_{\rm He~II})=276,114,43\times10^{50}$~photons~s$^{-1}$ for a $10^6M_\odot$ black hole. The Pop~III starburst spectrum is taken from \citet{tumlinson_nucleosynthesis_2004}, which gives $(Q_{\rm H~I},Q_{\rm He~I},Q_{\rm He~II})=38.10,22.97,1.30\times10^{50}$~photons~s$^{-1}$ for a $10^4M_\odot$ burst and  $(Q_{\rm H~I},Q_{\rm He~I},Q_{\rm He~II})=3810,2297,130\times10^{50}$~photons~s$^{-1}$ for a $10^6M_\odot$ burst. 

In some calculations, we place these sources in a dark matter halo of mass $10^8 M_\odot$ at $z=10$. Such a halo will contain of order $10^7 M_\odot$ of baryonic material. If this halo were able to convert baryons into stars as efficiently as the most efficient halos at $z=0$ \citep{leauthaud_new_2012}, this would imply a total stellar mass of order $10^6 M_\odot$, equal to our more massive starburst model. In reality, the efficiency of both black hole and star formation is likely much lower in such a halo (due to feedback effects), so we additionally explore $10^4 M_\odot$  starburst and AGN models in such a halo. Even this is arguably a very high mass for a black hole at such redshifts in a $10^8M_\odot$ halo. As such, our derived escape fractions are likely to be optimistic.

\begin{figure}
 \begin{center}
 \begin{tabular}{cc}
 DS94 profile; $10^4M_\odot$ \AGN; X-rays included &  DS94 profile; $10^4M_\odot$ \AGN; X-rays excluded \\
 \includegraphics[width=80mm]{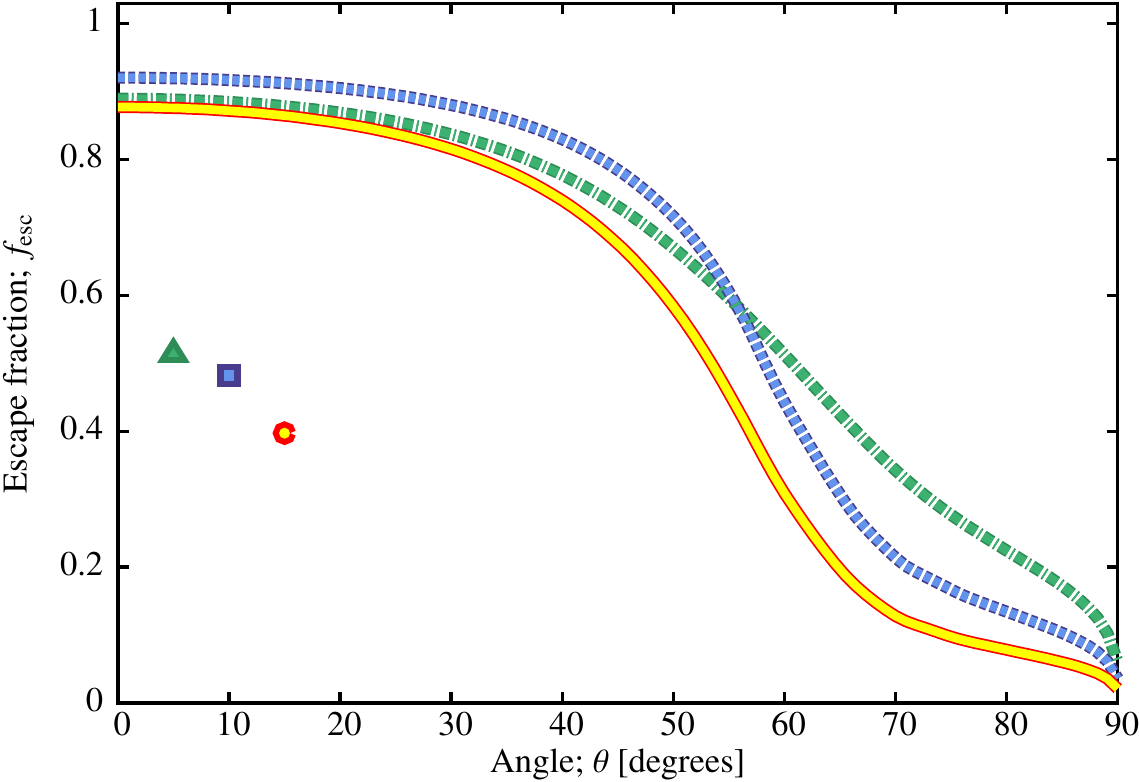} &
 \includegraphics[width=80mm]{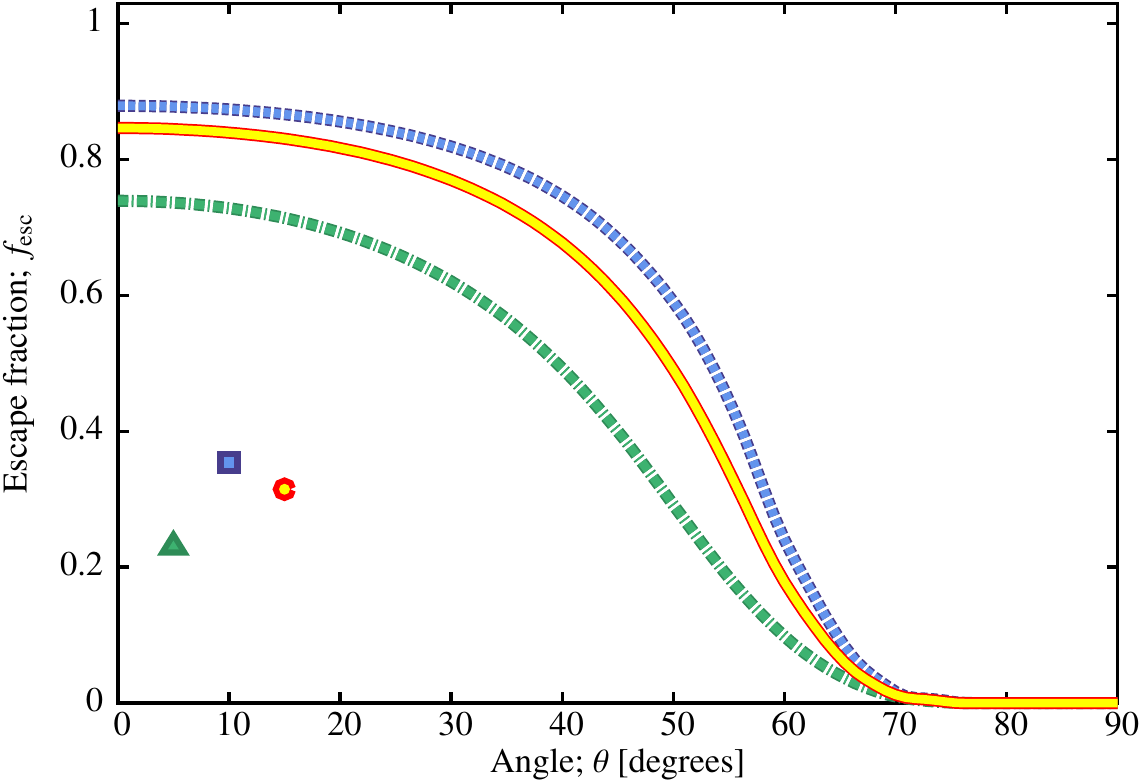}\vspace{-30mm}\\
\hspace{-10mm}\includegraphics[width=20mm]{legend3.pdf} \\
 \end{tabular}
 \end{center}
 \vspace{+10mm}\caption{\emph{Left panel:} The escape fraction as a function of angle, $\theta$, averaged over all realizations of the \protect\citet{dove_photoionization_1994} Gaussian model for a $10^4M_\odot$ \protect\AGN\ with clouds of radius 30~pc, filling factor 20\% and cloud overdensity factor of $C=10$. Same notation as Figure~\ref{fig:DS:single}. \emph{Right panel:} The result for the same model but with X-rays removed from the \protect\SED.}
 \label{fig:6}
\end{figure}

We highlight key trends in the escape fractions as a function of angle for a range of source masses, and the presence of X-rays. These trends, as shown in Figure~\ref{fig:6} and Table~\ref{table:1}, include:
\begin{itemize}

\item The impact of AGN and Pop~III sources is quite different, and the effects of switching off X-rays in these two cases differ
significantly, because the AGN are  more X-ray dominated;

\item The difference between a blackbody spectrum (as used in \S\ref{sec:DS}) and these sources is substantial, owing to the $Q$-values being significantly higher (lower) for the blackbody spectrum relative to the AGN (Pop~III starburst) -- compare, e.g., the $Q$ values  for the $10^4$  M$_\odot$ AGN or stellar cluster with the relevant values in Section 4.1.

\item Once the source mass exceeds about $10^5$ M$_\odot$, the background gas \citep{dove_photoionization_1994} ``saturates" and the escape fraction approaches unity for H and He;

\item Interestingly, X-rays boost the escape fractions overall, and are particularly significant for the lower-mass AGN or starburst masses. We also see clearly the advancing of the He~III I-front relative to that of H~II or He~II, when X-rays are present.
\end{itemize}

In summary, the values of $f_{\rm esc}$ found in the above subsections using the background density from \citet{dove_photoionization_1994} are quite high, more than
80--90\% in some cases for H~I and He~I. These cases are meant to be illustrative rather than exact as we are placing a variety of sources (including primordial stars and quasars) in what is essentially a Milky Way-type disk. Some of these sources also have significantly higher $Q$-values than that assumed for a single O-type star, $10^{49}$ s$^{-1}$, in \citet{dove_photoionization_1994}. In the following subsection we will explore primordial sources embedded in dark matter halos appropriate to the early Universe.

%
%

\subsection{Variations in Galaxy Profile and/or Source Location}\label{sec:profile}

Finally, we consider two alternatives to the \citet{dove_photoionization_1994} galaxy profile: the standard \NFW\ profile (\citealt{navarro_universal_1997}; with the gas profile appropriately scaled to the dark matter density through cosmological parameters), and the $\beta$-profile which we use to approximate the density distribution of gas in hydrostatic equilibrium at the virial temperature in an \NFW\ halo \citep{ricotti_feedback_2000}. For the \NFW\ profile, the density normalization assumes that the halo has a virial density contrast as predicted by the spherical top-hat collapse model for our chosen cosmology ($\Omega_{\rm M}=0.275$, $\Omega_\Lambda=0.725$; \citealt{komatsu_seven-year_2011}), and the scale radius is set using the fitting function of \citet{gao_redshift_2008}. For the $\beta$-profile, we set parameters using the results of \citet{ricotti_feedback_2000} such that the profile approximates that of gas in hydrostatic equilibrium in an \NFW\ density profile. 

We explored cases for a range of galaxy dark matter masses but display here the results for a $10^8$ $M_\odot$ halo, whose virial radius is about 1.4 (physical) kpc at $z=10$. In the \citet{dove_photoionization_1994} model used in earlier sections, the density profile drops away from the source as a Gaussian; thus, the escape fraction quickly reaches a constant value with distance from the source. In the case of an \NFW\ or $\beta$-profile, however, the density falls off only as $r^{-3}$ on large scales, so the escape fraction converges more slowly. We therefore use the virial radius of the halo as the radius at which to stop integrating and measure the escape fraction. The virial radius marks the dividing line between the galactic environment and the circumgalactic medium (CGM). Therefore, any photons escaping beyond the virial radius are able to ionize the CGM, in keeping with the usual definition of the escape fraction.

For these two density profiles (\NFW\ and $\beta$), we consider cases where the source is at the center, and cases where the source is placed 100~pc from the halo center ($\sim$ 10\% of the halo virial radius). The figures displayed in this section show the escape fraction at the virial radius as a function of angle, as in earlier sections of the paper. For the off-center source cases, note that $\theta=-90^\circ$ ($\theta=+90^\circ$) is the direction from the source directly through (directly away from) the center of the halo, and, consequently, always has the lowest (highest) escape fraction. The angle $\theta = 0^\circ$ corresponds to directions perpendicular to the halo center from the source. Most of the solid angle is around $\theta = 0^\circ$, so this typically gives the best estimate of the mean escape fraction.

\begin{figure}
 \begin{center}
 \begin{tabular}{cc}
 $\beta$-profile; $10^4M_\odot$ \AGN; X-rays included &  $\beta$-profile; $10^4M_\odot$ \AGN; X-rays excluded \\
 \includegraphics[width=80mm]{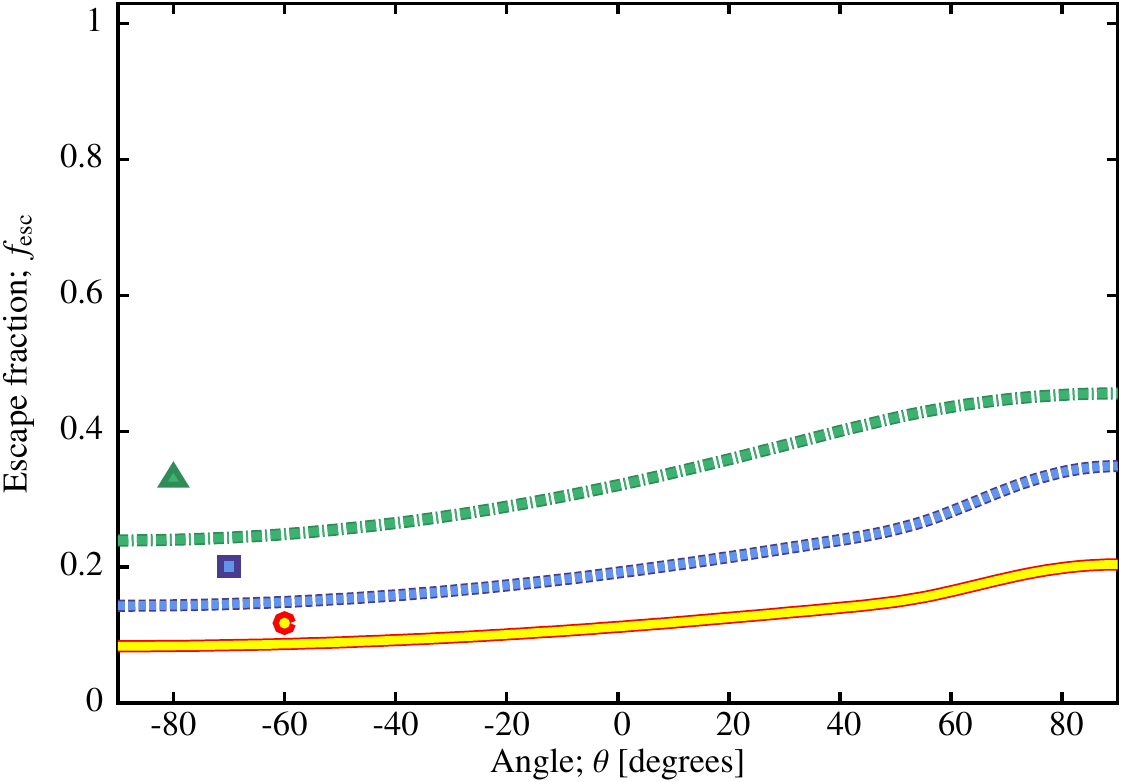} &
 \includegraphics[width=80mm]{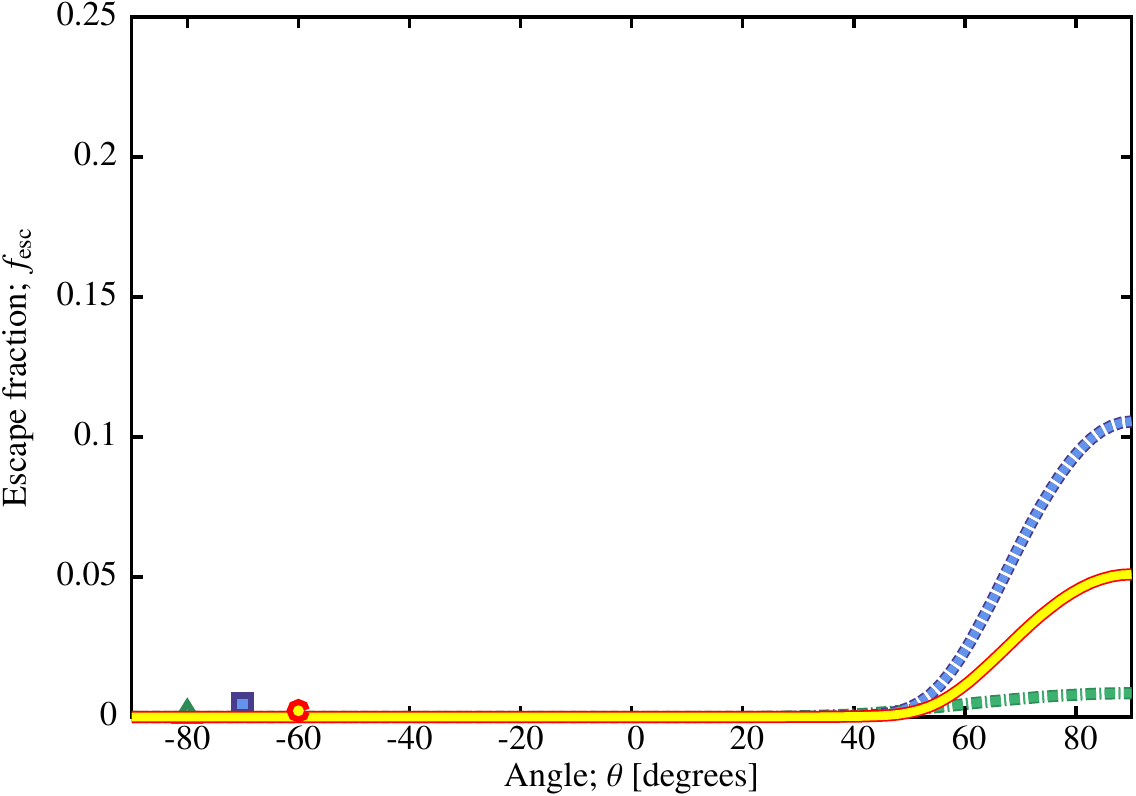}\vspace{-50mm}\\
\hspace{-10mm}\includegraphics[width=20mm]{legend3.pdf} \\
 \end{tabular}
 \end{center}
 \vspace{+30mm}\caption{\emph{Left panel:} The escape fraction as a function of angle, $\theta$, averaged over all realizations of a $\beta$-profile appropriate to a $10^8M_\odot$ \protect\NFW\ halo at $z=10$ for a $10^4M_\odot$ \protect\AGN\ placed 100~pc from the halo center and with clouds of radius 30~pc, filling factor 20\% and cloud overdensity factor of $C=10$. Same notation as Figure~\ref{fig:DS:single}. \emph{Right panel:} The result for the same model but with X-rays removed from the \protect\SED.}
\label{fig:betaAgn1e4}
\end{figure}

\begin{figure}
 \begin{center}
 \begin{tabular}{cc}
$10^4M_\odot$ Pop~III; $\beta$-profile & $10^4M_\odot$ Pop~III; \NFW\ profile \\
 \includegraphics[width=80mm]{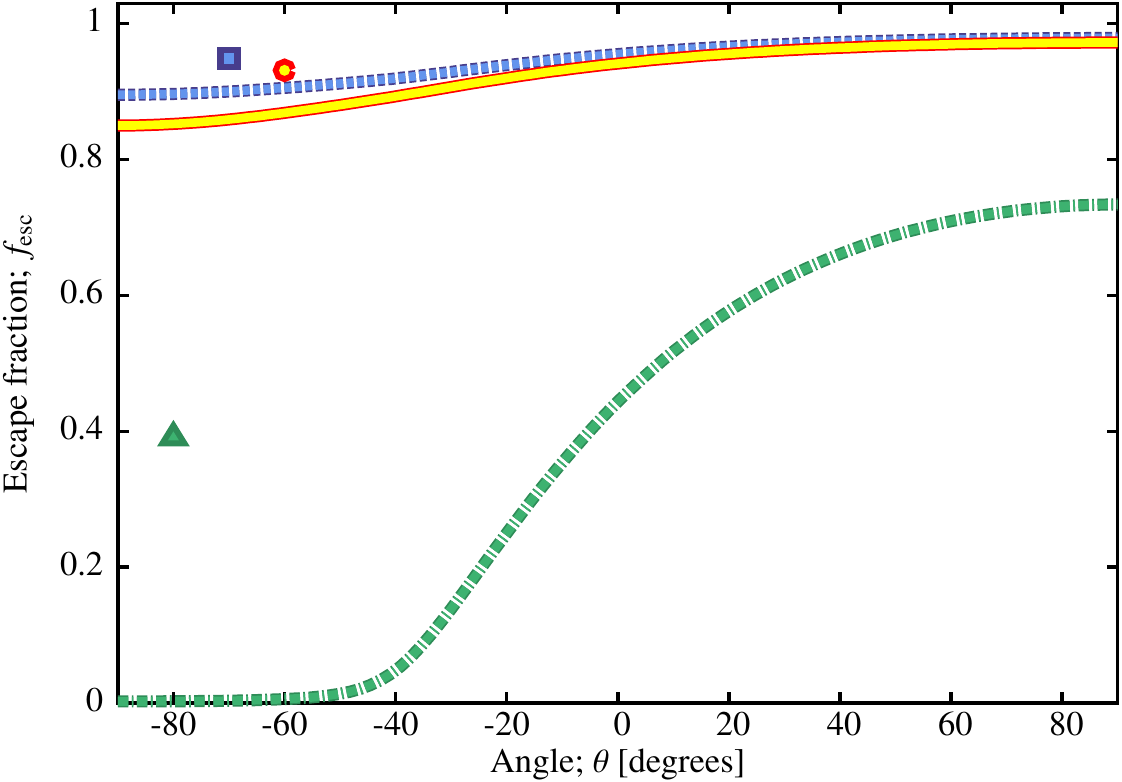} &
 \includegraphics[width=80mm]{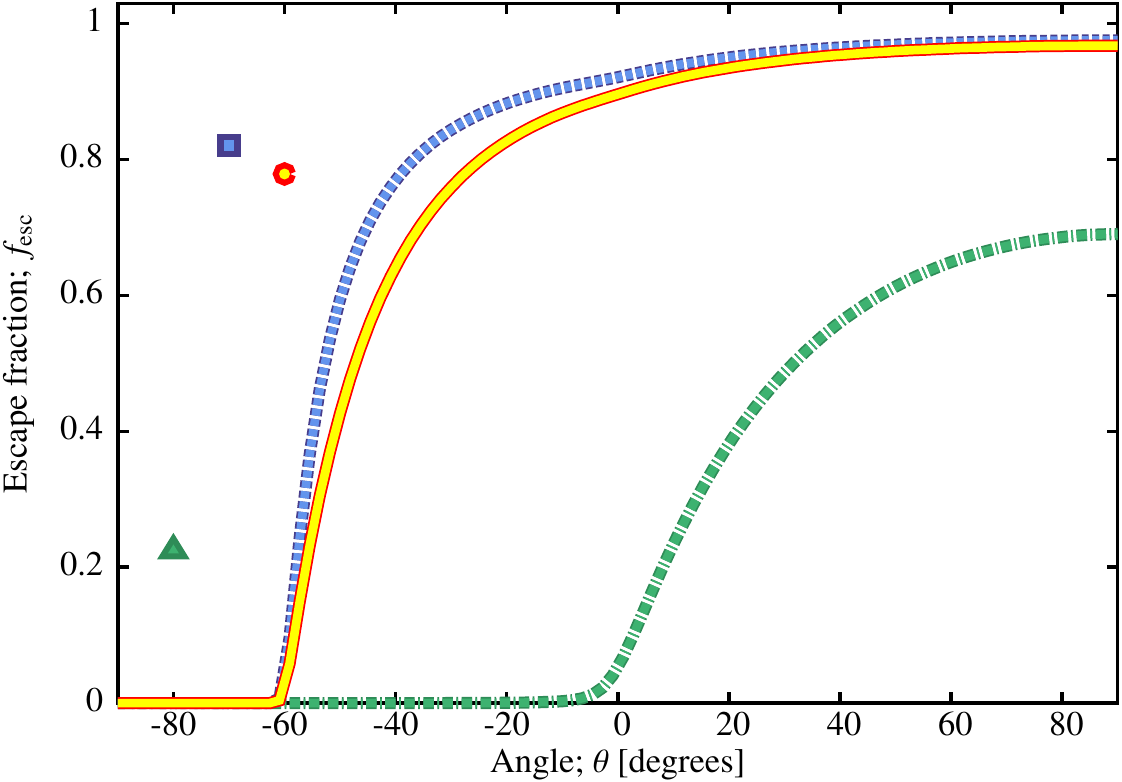}\vspace{-30mm}\\
\hspace{+30mm}\includegraphics[width=20mm]{legend3.pdf} \\
 \end{tabular}
 \end{center}
 \vspace{+10mm}\caption{\emph{Left panel:} The escape fraction as a function of angle, $\theta$, averaged over all realizations of a $\beta$-profile appropriate to a $10^8M_\odot$ \protect\NFW\ halo at $z=10$ for a $10^4M_\odot$ Pop~III cluster placed 100~pc from the halo center and with clouds of radius 30~pc, filling factor 20\% and cloud overdensity factor of $C=10$. Same notation as Figure~\ref{fig:DS:single}. \emph{Right panel:} The result for the same model but using the \protect\NFW\ itself rather than the corresponding $\beta$-profile.}
\label{fig:PopIII1e4}
\end{figure}

\begin{figure}
 \begin{center}
 \begin{tabular}{cc}
 $\beta$-profile; $10^4M_\odot$ \AGN; X-rays included & $\beta$-profile; $10^4M_\odot$ \AGN; X-rays excluded \\
 \includegraphics[width=80mm]{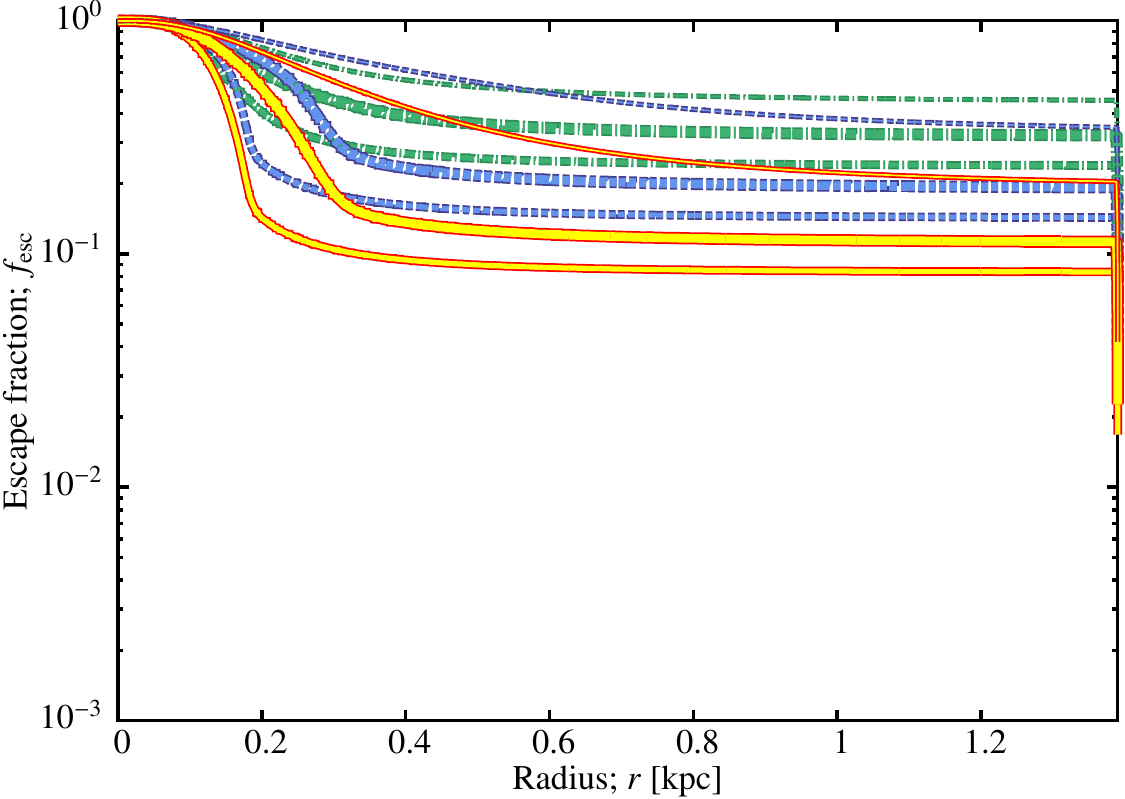} &
 \includegraphics[width=80mm]{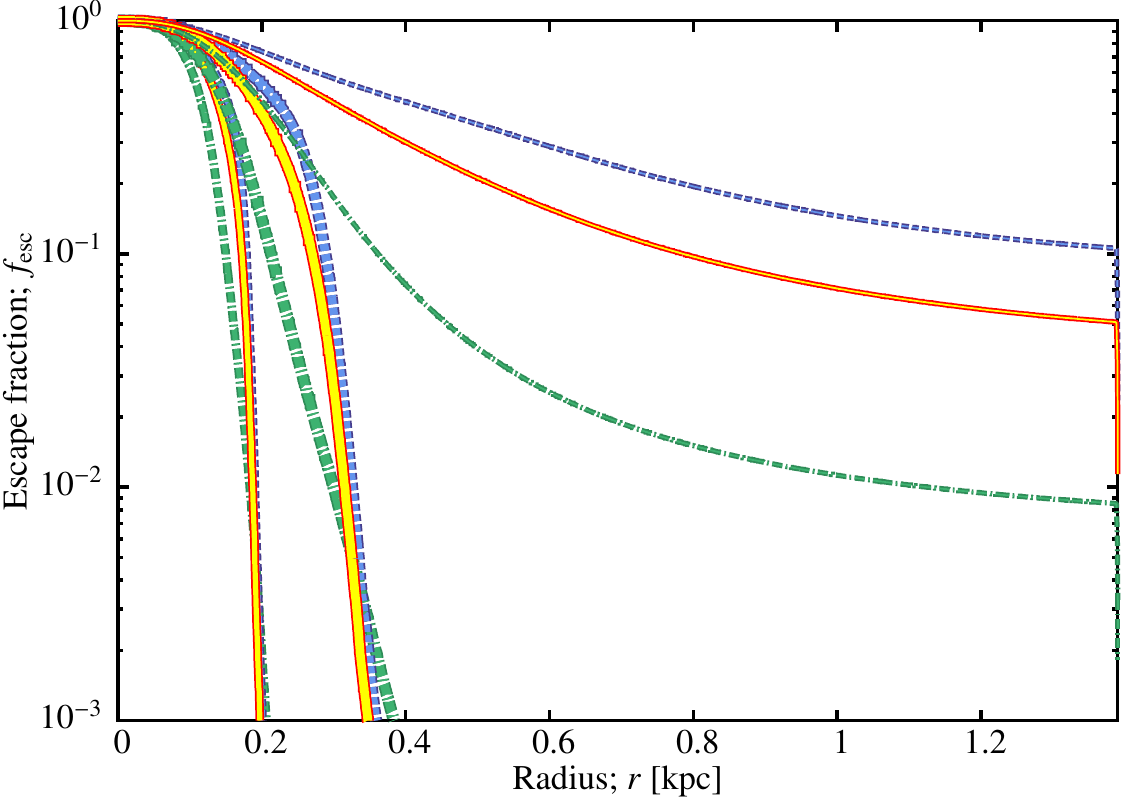}\vspace{-30mm}\\
\hspace{+30mm}\includegraphics[width=20mm]{legend3.pdf} \\
 \end{tabular}
 \end{center}
 \vspace{+10mm}\caption{\emph{Left panel:} The escape fraction (thick lines) and ionization fraction (thin lines) as a function of distance, $r$, from the source, averaged over all realizations of a $\beta$-profile appropriate to a $10^8M_\odot$ \protect\NFW\ halo at $z=10$ for a $10^4M_\odot$ \protect\AGN\ placed 100~pc from the halo center and with clouds of radius 30~pc, filling factor 20\% and cloud overdensity factor of $C=10$. Same notation as Figure~\ref{fig:DS:single}. \emph{Right panel:} The result for the same model but with X-rays removed from the \protect\SED.}
 \label{fig:betaRadial}
\end{figure}

Considering first the case of a $\beta$-profile with a $10^4M_\odot$ mass \AGN\ black-hole source (Figure~\ref{fig:betaAgn1e4}), there is a large difference when X-rays are excluded from the \AGN\ \SED, indicating that it is primarily the X-rays that are able to penetrate the halo and escape into the \IGM. Furthermore, in the case including X-rays, it is the He~II continuum photons (energies $\geq$ 4 Ryd) which have the highest escape fraction, due to the low cross-section for photons at these energies.
For the $10^4$ M$_\odot$ Pop~III starburst (Figure  \ref{fig:PopIII1e4}), $f_{\rm esc}$ is nearly unity for H~I and He~II ionizing radiation for all of the escape directions for a $\beta$-profile. The $f_{\rm esc}$ values for He~II ionizing radiation is significantly lower for all angles, approaching zero towards the direction of the galaxy center. If the source is instead placed precisely at the halo center, the escape fraction (which is now independent of angle due to the spherical symmetry) corresponds to that at $\theta=-15^\circ$ in Figure~\ref{fig:PopIII1e4}.

In the cases of the $10^6$ M$_\odot$ AGN, the values of $f_{\rm esc}$ saturate quickly to $1$ for all species, except for the \NFW\ cases near the very core (owing to the high densities there). X-rays make a difference in the direction of center but not otherwise. Hence, we do not display them, although here too, the He curve gets ahead of H in the very core of the galaxy. We find similar results for the case of the $10^6$ M$_\odot$ Pop~III starburst, except that there is very little difference with X-rays included owing to the low X-ray contribution to the \SED\ in this case.

Figure~\ref{fig:betaRadial} shows the dependence of escape fraction on radial distance for a $10^4M_\odot$ \AGN\ source placed 100~pc from the center of our $\beta$-profile model. The radial dependence is shown along three different directions---directly away from the halo center, perpendicular to the direction to the halo center, and directly through the halo center. In the left panel, which includes X-rays, the escape fractions quickly asymptote to almost constant values of between 10\% and 60\% in all directions. The right panel of Figure~\ref{fig:betaRadial} shows the same calculation but with X-rays removed from the source \SED. The behavior is substantially different in this case. In directions through the halo center and perpendicular to that direction the escape fractions plummet rapidly, indicating that it was the X-rays which were able to escape through these directions most easily. Directly away from the center, the escape fractions remain significant, but continue to fall out to the halo's virial radius.

In summary for this subsection, we find that  an \NFW\ gas distribution produces much lower escape fractions overall for both H and He relative to the $\beta$-profile, particularly through the halo center where the density rises rapidly. Additionally, the presence of X-rays boosts $f_{\rm esc}$ for H~I, He~I and He~II; this is especially dramatic for He~II.

\section{CONCLUSIONS AND DISCUSSION}

We have calculated the escape fraction of ionizing radiation, $f_{\rm esc}$, for both hydrogen and helium using 1D semi-analytic models of galaxies ranging from those typical of the early universe to disk-type galaxies that are not heavily dust-obscured. We have considered many scenarios that vary  the galaxy density profile, the source type (first stars or AGN), location, and spectral index,  as well as cloud overdenisty factors.  Key new points in our work include calculations of $f_{\rm esc}$ for He~I and He~II ionizing radiation, and the inclusion of the process of partial gas ionization by X-rays. We find that X-rays play an important role in boosting the $f_{\rm esc}$ values for H~I, He~I and He~II, especially He~II. Escape fractions from primordial halos containing a bright \AGN\ may reach as high as 30:20:10\% for H~II:He~II:He~III (mostly due to the significant contribution from X-rays in such sources), while a burst of Pop.~III star formation could result in escape fractions as high as 90:90:40\% for H~II:He~II:He~III. For sufficiently hard first-light sources, the helium ionization fronts closely track or even advance beyond that of hydrogen.

These calculations have an impact on a number of different observational probes of first-light sources and reionization models. First, the current WMAP9 results indicate that the Thomson optical depth, once covariances are accounted for, is $\tau_e \sim$ 0.08--0.09 \citep{2012arXiv1212.5226H}, allowing for an extended period of partial ionization of H and /or He beyond $z \sim$ 6. As discussed in \citet{Shull:08}, the need for an extreme reionization scenario is reduced for H, if $f_{\rm esc, He}$ is competitive with $f_{\rm esc, H}$. With primordial helium at $Y = 0.25$ by mass, the number ratio $y = n_{\rm He}/n_{\rm H} = 1/12$.  The inclusion of helium in models typically increases $\tau_e$ by 16\% relative to that for hydrogen only \citep{Shull:08}, accounting for singly ionized helium at $3 \leq z \leq 7$ and doubly ionized helium at $z < 3$.

Second, the values of $f_{\rm esc}$ and its redshift evolution can impact the evolution of the optical depth of He~II and the ratio of He~III to He~II at $z \sim 2.5$--3 \citep{khaire_he_2012, shull_hst-cos_2012}, during late He~II reionization. The ratio of \ion{He}{2} to \ion{H}{1} column densities, denoted by $\eta$, has been observed to have factor-of-ten variations on Mpc scales, along the sightlines to several ``He-II quasars" at $z \sim$ 2.4--2.9 \citep{shull_fluctuating_2004,shull_hst/cos_2010,fechner_uv_2006,syphers_hst/cos_2012}.  Because \ion{He}{2} is photoionized by 4-Ryd continuum radiation, these $\eta$-variations are probably dominated by fluctuations in the intrinsic spectra of the AGN sources, modulated by \ion{He}{2} opacity changes in the IGM (primarily from Lyman-limit systems). Ultraviolet spectroscopic studies of low-redshift AGN \citep{zheng_composite_1997,shull_hst-cos_2012} find rest-frame (1--2 Ryd) ionizing spectra, $F_{\nu} \propto \nu^{-\alpha}$, with a wide range of spectral indices from hard ($\alpha \approx 0.5$) to soft ($\alpha \approx 3$). This is consistent with the range of the critical spectral index of $\alpha \sim$ 0.5--1.8 derived in Section 3 when accounting for frequency-dependent effects in the source spectrum. The composite EUV spectra of low-redshift AGN have recently been measured to have $\langle \alpha \rangle \approx 1.4-1.5$ between 1.0--1.5~Ryd  \citep{shull_hst-cos_2012}.

Last, calculations of $f_{\rm esc}$ impact the converse problem of radiation trapped in primordial galaxies, and have consequences for detecting these   galaxies through emission-line signatures. Such emission-line signatures of very hot, low metallicity massive stars \citep{tumlinson_zero-metallicity_2000,tumlinson01} include strong \ion{He}{2} recombination lines such as $\lambda 1640$ and $\lambda4686$.

Looking ahead, science drivers for future UV spectroscopic missions \citep{Tumlinson:12,McCandliss:12} include telescopes and spectrographs with spectral coverage down to the hydrogen Lyman limit (912~\AA).  This coverage will allow one to directly measure escaping (hydrogen) LyC radiation from low-redshift starburst galaxies.  This far-UV band also probes \ion{He}{2} Gunn-Peterson absorption between $z \approx 2-3$, an interval that spans the patchy  \ion{He}{2} post-reionization epoch in the rest-frame of the intervening IGM \citep{shull_fluctuating_2004,shull_hst/cos_2010,Syphers:13}.  Observations of ``He~II quasars" at somewhat higher redshifts, $z = 3.0-3.5$ 
     \citep{syphers_high_2009,syphers_ten_2009,syphers_hst/cos_2012,worseck_end_2011} can observe the \ion{He}{2} absorption during the epochs when the \ion{He}{3} ionized bubbles have not yet overlapped. Placing direct limits on $f_{\rm esc, He}$  faces significant observational challenges, however, and the likelihood of a FUV instrument as a successor to COS remains uncertain at present.

\acknowledgements

We thank Jason Tumlinson for useful correspondence on the spectra from metal-free stellar populations, and Yue Zhao, Yang Wang, and Long Yan Yung for useful input. AV gratefully acknowledges support from Research Corporation through the Single Investigator Cottrell College Science Award, and from the University of San Francisco Faculty Development Fund. JMS acknowledges support from NASA through Astrophysics Theory grant NNX07-AG77G.

\bibliographystyle{apj}
\bibliography{fesc}

\end{document}